\documentclass[12pt]{iopart}

\usepackage{iopams}  
\usepackage{graphicx}
\usepackage{subfig}
\usepackage{xcolor}
\usepackage{ulem}
\usepackage{adjustbox}
\usepackage{pdflscape}

\usepackage{multirow}

\newcommand{\bq}{\begin{equation}}
\newcommand{\eq}{\end{equation}}
\newcommand{\bqn}{\begin{eqnarray}}
\newcommand{\eqn}{\end{eqnarray}}
\newcommand{\nb}{\nonumber}
\newcommand{\lb}{\label}

\begin{document}

\title{Thermodynamics of Einstein-Aether Black Holes}

\author{R. Chan$^{1}$, M. F. A. da Silva$^{2}$ and V. H. Satheeshkumar$^{3}$} 

\address{	
			$^{1}$Coordena\c{c}\~{a}o de Astronomia e Astrof\'{i}sica,
			Observat\'{o}rio Nacional (ON), Rio de Janeiro, RJ 20921-400, Brazil
			\\
			$^{2}$Departamento de F\'{i}sica Te\'{o}rica, 
			Universidade do Estado do Rio de Janeiro (UERJ), Rio de Janeiro, RJ 20550-900, Brazil
			\\
			$^{3}$Departamento de F\'{\i}sica, Universidade Federal do Estado do Rio de Janeiro (UNIRIO), Rio de Janeiro, RJ 22290-240, Brazil
		}
			
\ead{chan@on.br, mfasnic@gmail.com, vhsatheeshkumar@gmail.com}
	
\begin{abstract}
We analyze several spherically symmetric exterior vacuum solutions allowed by the Einstein-Aether (EA) theory with a non static aether and study the thermodynamics of their Killing and universal horizons. We show that there are five classes of solutions corresponding to different values of a combination of the free parameters, $c_{2}$, $c_{13}=c_1+c_3$ and $c_{14}=c_1+c_4$, which are: \textcolor{black}{(A) $c_2 \ne 0$ and $c_{13} \ne 0$ and $c_{14}=0$, (B) $c_2 \ne 0$ and $c_{13} = 0$ and $c_{14} = 0$, (C) $c_2 = 0$ and $c_{13} \ne 0$ and $c_{14} = 0$, (D) $c_2 = 0$,  $c_{13} = 0$ and $c_{14} \ne 0$, and (E) $c_2 = - c_{13} \ne 0$ and $c_{14} \ne 0$.}
We present explicit analytical solutions for these five cases. All these cases have singularities at $r=0$ and are asymptotically flat spacetimes and possess both Killing and universal horizons with the universal horizons always being inside the Killing horizons. Finally, we compute the surface gravity, the temperature, the entropy and the first law of thermodynamics for the  universal horizons.
\end{abstract}

\section{Introduction}

The two recurring themes in our quest for a theory of quantum gravity have been to either introduce a new fundamental symmetry such as supersymmetry or break a fundamental symmetry such as Lorentz invariance (LI). The breaking of LI is found to make the construction of quantum gravity a possible task, at least on paper, see \cite{Li:2015itk} for an example. Currently, one of the principal guiding lights in quantum gravity research is the study of black hole thermodynamics. Thus, it is interesting to investigate the thermodynamics of black holes in a gravitational theory that explicitly breaks LI. This is precisely what we intend to do in this work. 

The LI is an exact symmetry in special relativity, quantum field theories, and the standard model of particle physics, while in General Relativity (GR) it is only a local symmetry in freely falling inertial frames  \cite{Moore:2013sra}. The violation of LI in the gravitational sector is not as well explored as in matter interactions where it is highly constrained by several precision experiments \cite{Bars:2019lek}. Jacobson and his collaborators introduced and analyzed a general class of vector-tensor theories called the Einstein-Aether (EA) theory \cite{Jacobson:2000xp} \cite{Eling:2003rd} \cite{Jacobson:2004ts} \cite{Eling:2004dk} \cite{Foster:2005dk} to study the  effects of violation of LI in gravity. A brief review of the vector-tensor theories of gravity can be found in \cite{Satheeshkumar:2021zvl}. The first spherical static vacuum solutions in the EA theory were obtained by Eling and Jacobson in 2006 \cite{Eling2006}. Since then several more solutions have been found including our recent analytical solutions for static aether \cite{Chan2020}.  Most of the literature on black holes in EA theory can be found in the papers \cite{Eling2007}-\cite{Adam2021}.

The subject of black hole thermodynamics was born in 1974 when Stephen Hawking \cite{Hawking:1974rv} mathematically showed that a black hole
radiates as though it had a temperature proportional to its surface gravity, and he asserted that the similarities
 between the laws of black-hole mechanics \cite{Bardeen:1973gs} and the laws of thermodynamics
were more than a coincidence \cite{Bekenstein:1973ur}. Hawking
also obtained a precise relationship between the entropy of the black hole and its surface
area \cite{Hawking:1975vcx}. The reviews \cite{Carlip:2014pma} and \cite{Sarkar:2019xfd} give a good outline of the subject.

The paper is organized as follows.  Section $2$ briefly presents the EA theory, whose field equations are solved for a general spherically symmetric metric in Section $3$. In Section $4$, we present the basics of black hole thermodynamics. \textcolor{black}{In Sections $5$ -- $12$ we present the explicit analytical solutions and study their thermodynamics. We summarize our results in Section $13$. The field equations are quite long, so we have relegated them to the Appendix.} 

\section{Field equations in the EA theory }

The general action of the EA theory is given by  
\bq 
S = \int \sqrt{-g}~(L_{\rm Einstein}+L_{\rm aether}+L_{\rm matter}) d^{4}x,
\label{action}
\eq
where, the first term is the usual Einstein-Hilbert Lagrangian, defined by $R$, the Ricci scalar, and $G$, the EA gravitational constant, as
\bq 
L_{\rm Einstein} =  \frac{1}{16\pi G} R. 
\eq
The second term, the aether Lagrangian is given by
\bq 
L_{\rm aether} =  \frac{1}{16\pi G} [-K^{ab}{}_{mn} \nabla_a u^m
\nabla_b u^n +
\lambda(g_{ab}u^a u^b + 1)],
\lb{LEAG}
\eq
where the tensor ${K^{ab}}_{mn}$ is defined as
\bq 
{K^{ab}}_{mn} = c_1 g^{ab}g_{mn}+c_2\delta^{a}_{m} \delta^{b}_{n}
+c_3\delta^{a}_{n}\delta^{b}_{m}-c_4u^a u^b g_{mn},
\lb{Kab}
\eq
being the $c_i$ dimensionless coupling constants, and $\lambda$
a Lagrange multiplier enforcing the unit timelike constraint on the aether, and 
\bq
\delta^a_m \delta^b_n =g^{a\alpha}g_{\alpha m} g^{b\beta}g_{\beta n}.
\eq
Finally, the last term, $L_{\rm matter}$ is the matter Lagrangian, which depends on the metric tensor and the matter field.

In the weak-field, slow-motion limit EA theory reduces to Newtonian gravity with a value of  Newton's constant $G_{\rm N}$ related to the parameter $G$ in the action (\ref{action}) by  \cite{Garfinkle2007},
\bq
G = G_N\left(1-\frac{c_{14}}{2}\right).
\lb{Ge}
\eq
Here, the constant $c_{14}$ is defined as
\bq
c_{14}=c_1+c_4.
\lb{beta}
\eq

The field equations are obtained by extremizing the action with respect to independent  variables of the system. The variation with respect to the Lagrange multiplier $\lambda$ imposes the condition that $u^a$ is a unit timelike vector, thus 
\bq
g_{ab}u^a u^b = -1,
\label{LagMul}
\eq
while the variation of the action with respect $u^a$, leads to \cite{Garfinkle2007}
\bq
\nabla_a J^a_b + c_4 a_a \nabla_b u^a + \lambda u_b = 0,
\eq
where,
\bq
J^a_m=K^{ab}_{mn} \nabla_b u^n,
\eq
and
\bq
a_a=u^b \nabla_b u_a.
\lb{aa}
\eq
The variation of the action with respect to the metric $g_{mn}$ gives the dynamical equations,
\bq
G^{Einstein}_{ab} = T^{aether}_{ab} +8 \pi G  T^{matter}_{ab},
\label{EA}
\eq
where 
\bqn
G^{Einstein}_{ab} &=& R_{ab} - \frac{1}{2} g_{ab} R, \nb \\
T^{aether}_{ab}&=& \nabla_c [ J^c\;_{(a} u_{b)} + u^c J_{(ab)} - J_{(a} \;^c u_{b)}] - \frac{1}{2} g_{ab} J^c_d \nabla_c u^d+ \lambda u_a u_b  \nb \\
& & + c_1 [\nabla_a u_c \nabla_b u^c - \nabla^c u_a \nabla_c u_b] + c_4 a_a a_b, \nb \\
T^{matter}_{ab} &=&  \frac{- 2}{\sqrt{-g}} \frac{\delta \left( \sqrt{-g} L_{matter} \right)}{\delta g_{ab}}.
\label{fieldeqs}
\eqn

Later, when we solve the field equations (\ref{EA}), we do take into consideration the equations 
(\ref{LagMul})-(\ref{aa}) in the process of simplification. Thus, in this paper (as in the equations (\ref{Gtt})-(\ref{Gphiphi}) below) we seem to solve only the dynamical equations, but in fact, we are also solving the equations arising from the variations of the action with respect $\lambda$ and $u^a$.

In a more general situation, the Lagrangian of GR is recovered, if and only if, the 
coupling constants are identically zero, e.g., $c_1=c_2=c_3=c_4=0$, 
considering the equations
(\ref{Kab}) and (\ref{LagMul}).

\section{Spherical Solutions of EA field equations}

We start with the most general spherically symmetric static metric
\bq
ds^2= -A(r) dt^2+B(r) dr^2 +r^2 d\theta^2 +r^2 \sin^2 \theta d\phi^2.
\lb{ds2}
\eq
In accordance with equation (\ref{LagMul}), the aether field is assumed to be unitary and timelike, chosen as
\bq
u^a= \left[ \frac{b(r)}{\sqrt{-a(r)^2 B(r)+b(r)^2 A(r)}}, \frac{a(r)}{\sqrt{-a(r)^2 B(r)+b(r)^2 A(r)}}, 0, 0 \right],
\eq
where $x^\mu=(t,r,\theta,\phi)$ are the directions of the aether vector. Since the aether vector
is unitary we have $a(r)^2 B(r)-b(r)^2 A(r)=-1$. 
Substituting 
\bq
b(r)=\frac{\epsilon}{A(r)}\sqrt{A(r) [a(r)^2 B(r)+1]}
\lb{br}
\eq
from this last condition
we obtain the aether vector depending only of $a(r)$, where $\epsilon=\pm 1$.

\bq
u^a = \left[ \epsilon\frac{\sqrt{A(r) [a(r)^2 B(r)+1] }}{A(r)}, a(r), 0, 0\right],
\lb{ua}
\eq

The timelike Killing vector of the metric (\ref{ds2}) is giving by
\bq
{\chi}^{\alpha} = (-1, 0, 0, 0).
\lb{chia}
\eq
The Killing and the universal horizon \cite{Wang} \cite{Berglund2012} are obtained 
finding the largest root of
\bq
{\chi}^{\alpha} {\chi}_{\alpha} = 0,
\eq
and
\bq
{\chi}^{\alpha} {u}_{\alpha} = 0,
\eq
respectively, where ${\chi}^{\alpha}$ is the timelike Killing vector. In our case, 
\bq
{\chi}^{\alpha} {\chi}_{\alpha} =-A(r),
\lb{r_{kh}}
\eq
\bq
{\chi}^{\alpha} {u}_{\alpha} = \frac{A(r) b(r)}{\sqrt{-a(r)^2 B(r)+b(r)^2 A(r)}}.
\lb{r_{uh}}
\eq

In order to identify eventual singularities in the solutions, it is useful to calculate the Kretschmann scalar invariant K. For the metric (\ref{ds2}), it is  given by
\bqn
K &=&  \frac{1}{4r^4 B^4 A^4} \left( 16 B^4 A^4-32 B^3 A^4+16 B^2 A^4+8 A'^2 r^2 B^2 A^2+\right.\nb\\
&&\left. 8 B'^2 r^2 A^4+4 r^4 A''^2 B^2 A^2-4 r^4 A'' B A^2 B' A'-4 r^4 A'' B^2 A A'^2+\right.\nb\\
&&\left. r^4 B'^2 A'^2 A^2+2 r^4 B' A'^3 A B+r^4 A'^4 B^2 \right).
\lb{K}
\eqn

\textcolor{black}{
The field equations are given in full detail in the Appendices A 
(collecting the terms $c_2$, $c_{13}$ and $c_{14}$) and B 
(collecting the terms $c_{123}$, $c_{14}$ and $c_{2}$). 
Notice that assuming $c_{123}=0$ does not give the same field equations
that assuming $c_2+c_{13}=0$ or $c_2=0$ with $c_{13}=0$. 
This characteristic of the field equations imposes
different solutions for each case that will be clearer below.}

\textcolor{black}{
Solving simultaneously equations (\ref{Gtt})-(\ref{Gphiphi}) using Maple 16 we get five 
particular families of analytic solutions: (A) $c_2 \ne 0$ and $c_{13} \ne 0$ and $c_{14}=0$, (B) $c_2 \ne 0$ and $c_{13} = 0$ and $c_{14} = 0$, (C) $c_2 = 0$ and $c_{13} \ne 0$ and $c_{14} = 0$, (D) $c_2 = 0$,  $c_{13} = 0$ and $c_{14} \ne 0$, and (E) $c_2 = - c_{13} \ne 0$ and $c_{14} \ne 0$. Let us now analyze these five possible cases in detail. The solutions for only $c_2=0$ or only $c_{13}=0$} were not shown because they are
a special case of static aether, imposing $a(r)=0$ (see our previous work \cite{Chan2020} for the static aether case).

It was shown \cite{Ding:2015kba} that Smarr formula and corresponding first law of black hole mechanics exists for ranges of the $c_i$'s, $0 \le c_{14} < 2$, $c_{13} < 1$ and $2 + c_{13} + 3 c_2 > 0$. All our solutions fall within this interval. Imposing these limits on our choices, we get $c_2 > -2/3$, $ -2 < c_{13} < 1$ and $0 < c_{14} < 2$.

\section{Black Hole Thermodynamics}

We briefly discuss the thermodynamics of black holes for the sake of completeness. The essential ingredient for this is the existence of horizons, the usual event horizon in the relativistic case and universal horizon in the Lorentz-violating theories like the one we are considering here.  For more discussion on black hole thermodynamics in EA theory, we refer the readers to references \cite{Foster2006}, \cite{Ding2016a} and \cite{Berglund:2012fk, Ding:2018whp, Ding:2020bwa}. Recently, the thermodynamics of black holes in Einstein-Aether-Maxwell theory was investigated by the phase-space solution method in  \cite{Ding:2020bwa}. 

\textcolor{black}{
Besides the horizons, the other important quantity in black hole thermodynamics is the surface gravity $\kappa$. Once we have these two quantities, we can interpret the $\kappa/{2 \pi}$ as 
the temperature of the black hole and $A/4$ as the entropy of the black hole where $A$ is the surface area of the horizon. The surface gravity for Killing and universal horizons is defined respectively as (see the equations (37) and (39) of the reference \cite{Ding:2020bwa}), 
\bqn
\kappa_{kh} &=& \frac{1}{2}\sqrt{- {\chi}^{a;b} {\chi}_{b;a}} \, |_{r=r_{kh}},\nb\\
\kappa_{uh} &=& \frac{1}{2} u^b (u^a {\chi}_{a})_{;b} \, |_{r=r_{uh}}, 
\lb{termk}
\eqn
where the symbol semicolon means covariant derivative.
Calculating these quantities we get
\bqn
\kappa_{kh}=\frac{1}{2}\sqrt {\frac {A'^{2}}{B A}}\, |_{r=r_{kh}},\nb\\
\kappa_{uh}=-\frac {\epsilon a \left( A' a^{2}B + A' +2A a B  A' +A a^{2}B'  \right)  }{4\sqrt {A  \left( a  ^{2}B +1 \right) }}\, |_{r=r_{uh}}.
\eqn
We would like to mention that since any constant multiple of a Killing vector is also another Killing vector, it does not uniquely specify the scaling of the surface gravity; it can be changed by a constant factor by rescaling the Killing vector. If the spacetime is asymptotically flat,
then one can normalize the Killing vector at the spatial infinity thereby obtaining a
unique value for the surface gravity -- which is what we have done here.}

\textcolor{black}{
For the universal horizon,  we can directly write down (see equation (63) of \cite{Ding:2020bwa})} the surface gravity, temperature, entropy and the first law for the universal horizon respectively as, 
\bqn
T_{uh} &=& \frac{\kappa_{uh}}{2 \pi},\nb\\
S_{uh} &=& \frac{ \pi r^{2}_{uh}}{ G}=\frac{{A_{uh}}}{4 G},\nb\\
\delta S_{uh} &=& \frac{\delta M}{T_{uh}},
\lb{termu}
\eqn
where $A_{uh}$  is the the area of universal horizon. It is clear from the equations above that they are exactly the same as in GR. However, for a non-static aether, one may have to redefine surface gravity because of the contribution of aether vector field \cite{Ding:2020bwa} \textcolor{black}{as we have in Eq.(\ref{termk}).}

In order to compare our results with the GR black we calculate the
thermodynamical quantities for $r_{kh}=r_{uh}=r_{GR}=2M$ or the Schwarzschild radius $r_s = \frac{2G_N M}{c^{2}} $ giving
\bqn
\kappa_{GR}= \frac{G_N M}{r_s^{2}} = \frac{c^{4}}{4 G_N M} = \frac{1}{4M}  ,\nb\\
T_{GR}= \frac {\hbar  \kappa_{GR} }{2\pi c k_B } = {\frac {\hbar  c^{3} }{8\pi k_B G_N   M }} = {\frac {1}{8\pi M}},\nb\\
S_{GR}=\frac{k_{B} c^{3} A_{GR} }{4 \hbar G_N} = \frac { 4 \pi k_{B} G_N M^{2} }{\hbar c} = 4\pi {M}^{2},\nb\\
A_{GR}=4 \pi r_s^{2} = \frac{16 \pi G_N^{2} M^{2} }{c^{4}} = 16\pi {M}^{2},\nb\\
\delta S_{GR} = \frac { 8 \pi k_{B} G_N M\, \delta M\,}{\hbar c} = 8\pi M\,\delta M.
\eqn
We are explicitly showing all the constants because the gravitational coupling constant is different in GR and EA. But, from now on, we work with $G_N=c=\hbar=k_B=1$.

\begin{table*}
\centering
\begin{minipage}{100 mm}
\caption{Summary of the Aether Conditions}
\label{table1}
\begin{tabular}{@{}|c|c|c|c|c|}
\hline
Solution & $c_2$ & $c_{13}$ & $c_{14}$ & $c_i$ \\
\hline
\textcolor{black}{$A$} &   &   & 0 & $c_{13}<1$ and $2+c_{13}+2c_2>0$ \\
\hline
\textcolor{black}{$B$} &   & 0 & 0 &  $c_2 > -2/3$ \\
\hline
\textcolor{black}{$C$} & 0 &   & 0 &  $-2 < c_{13} < 1$ \\
\hline
\textcolor{black}{$D$} & 0 & 0 &   &  $0 < c_{14} < 2$ \\
\hline
\textcolor{black}{$E$} &   &   &  & \textcolor{black}{$c_{2}=-c_{13} \neq 0$} \\
\hline
\hline
\end{tabular}

\medskip
Here $c_i$ (i=1..4) are the parameters of the aether field.
In order to have a theory that obeys the solar system tests,
we must have $0 \le c_{14} < 2$, $c_{13}<1$ and $2+c_{13}+2c_2>0$ \cite{Foster:2005dk}.
\end{minipage}
\end{table*}

\section{{\bf Solutions for case \textcolor{black}{(A)}:} $\bf c_{2} \neq 0$  and $\bf c_{13} \neq 0$ and $\bf c_{14} = 0$}
%
The solution of the field equations (\ref{Gtt})-(\ref{Gphiphi}) for 
$c_{14}=0$ we get, 
\bqn
&&A=G_1+\frac{G_2}{r}+G_3 r^2+\frac{G_4}{r^4},\nb\\
&&B=\frac{r^4}{r^4+G_2 r^3+G_4},\nb\\
\eqn
where $G_1$, $G_2$, $G_3$ and $G_4$ are arbitrary integration constants
we have chosen $G_1=1$ and $G_3=0$ in order to have
a flat spacetime at infinity and $G_2=-|G_2|$ in order to have a resemblance
with the Schwarzschild solution as in the GR and $|G_2|=2M$, where $M$ is the 
Schwarzschild mass. 
Note again as in the Case C, the term $r^{-4}$ it could be compared in GR to the 
Hartle and Thorne \cite{Hartle1968} solution of a slowly rotating deformed relativistic star,
assuming the quadrupole moment is zero. 
Thus, $A(r)$ and $B(r)$ can be rewritten as
\bqn
&&A=1-\frac{2M}{r}+\frac{G_4}{r^4},\nb\\
&&B=\frac{r^4}{r^4-2M r^3+G_4}.
\eqn

The solutions for $a(r)$ and $b(r)$ are
\bqn
&&a = -\frac{8\zeta}{r^2(24 c_2+8 c_{13}) c_{13}} \sqrt{-G_4 c_{13}(3c_2+c_{13})^2},\nb\\
&&b =\frac{\epsilon\; r^4}{r^4-2M r^3+G_4}\sqrt{\frac{r^4 c_{13}-2 c_{13} M r^3+c_{13} G_4-G_4}{r^4 c_{13}}},
\eqn
where $-G_4 c_{13}(3c_2+c_{13})^2>0$ and
$({r^4 c_{13}-2 c_{13} M r^3+c_{13} G_4-G_4})/{c_{13}} > 0$, in order to ensure that the components of the aether vector are real. The first condition imposes that $G_4$ and $c_{13}$ must have opposite signs, while the second one is identical to the Case C, changing $E_4$ by $G_4$. Then, we have the same limit for $G_4$ in order to ensure that $b$ be real, that is,
\bq 
G_4 \ge  \frac{27}{16} \left( \frac{c_{13}}{c_{13}-1} \right) M^4.
\eq
Note that the solutions presented in this section depend explicitly on the parameters 
$c_{2}$ and $c_{13}$.

The Kretschmann scalar is given by
\bq
K = \frac{12 (4 M^2 r^6 -20 M G_4 r^3 +39 G_4^2)}{r^{12}}.
\eq
Note that $r=0$ is the singularity of the spacetime.

The Killing horizon equation is given by
\bqn
&&{\chi}^{\alpha} {\chi}_{\alpha} = -1+\frac{2M}{r}-\frac{G_4}{r^4}=0,\nb\\
\eqn
whose roots are
\bqn
&&r_{kh1,2,3,4} = \frac{1}{2} M+\eta_1 \frac{\sqrt{3}}{12}  \delta_2+\nb\\
&&\zeta \frac{1}{12} \sqrt{-\frac{-72 M^2 \delta_1 \delta_2+6 \delta_2 \delta_1^2+
288 \delta_2 G_4-144\;\sqrt{3} M^3  \delta_1}{\delta_1 \delta_2}},
\eqn
where 
\bqn
&&\delta_1=\left(432 G_4 M^2+12 \sqrt{-768 G_4^3+1296 G_4^2 M^4}\right)^\frac{1}{3},\nb\\
&&\delta_2=\sqrt{\frac{12 M^2 \delta_1+2 \delta_1^2+96 G_4}{\delta_1}}.
\eqn

The universal horizon equation is
\bqn
&&{\chi}^{\alpha} {u}_{\alpha} = \epsilon \sqrt{\frac{r^4 c_{13}-2 c_{13} M r^3+c_{13} G_4-G_4}{r^4 c_{13}}}, 
\lb{uhD}
\eqn
whose four solutions are
\bqn
&&r_{uh1,2,3,4} = \frac{1}{2} M+\eta_1 \frac{\sqrt{3}}{12}  \lambda_3+\nb\\
&&\zeta \frac{1}{12} \left[ -(-72 M^2 c_{13} \lambda_2 \lambda_3+
6\;.\;12^\frac{1}{3} \lambda_3  \lambda_2^2-24\;.\;12^\frac{2}{3} \lambda_3 G_4  c_{13}+\right.\nb\\
&&\left. 24\;.\;12^\frac{2}{3} \lambda_3 G_4  c_{13}^2-144 \sqrt{3} M^3  c_{13} \lambda_2)(\lambda_2 \lambda_3)^{-1} \right]^{\frac{1}{2}},
\eqn
where
\bqn
&&\lambda_1= 36 M^2+\sqrt{3} \sqrt{\frac{-256 c_{13} G_4+256 G_4+432 M^4 c_{13}}{c_{13}}},\nb\\
&&\lambda_2=\left[(-1+c_{13}) G_4  \lambda_1 c_{13}^2\right]^\frac{1}{3},\nb\\
&&\lambda_3=\sqrt{ \frac{12 M^2 c_{13} \lambda_2+2 12^\frac{1}{3} \lambda_2^2+8 G_4 12^\frac{2}{3} c_{13}^2-8 G_4 12^\frac{2}{3} c_{13}}{c_{13} \lambda_2}}.
\eqn

In order to have a real aether vector we must impose that
\bq
c_{13} < 1,
\eq
\bq 
c_{13}<-2(1+c_2),
\eq
and
\bq 
c_{13} G_4 \le 0.
\eq

In order to have real Killing horizons we must impose that
\bq
G_4 \le  \frac{27}{16} M^4,
\eq
and real universal horizons we must have
\bq 
G_4 \ge  \frac{27}{16} \left( \frac{c_{13}}{c_{13}-1} \right) M^4.
\eq
From these two conditions we have that
\bq 
G_4 =  \frac{27}{16} \left( \frac{c_{13}}{c_{13}-1} \right) M^4.
\lb{G4a}
\eq

We can observe that the condition (\ref{G4a})
satisfies simultaneously the conditions for a real aether vector.

\begin{figure}[!ht]
\centering
	\includegraphics[width=7cm]{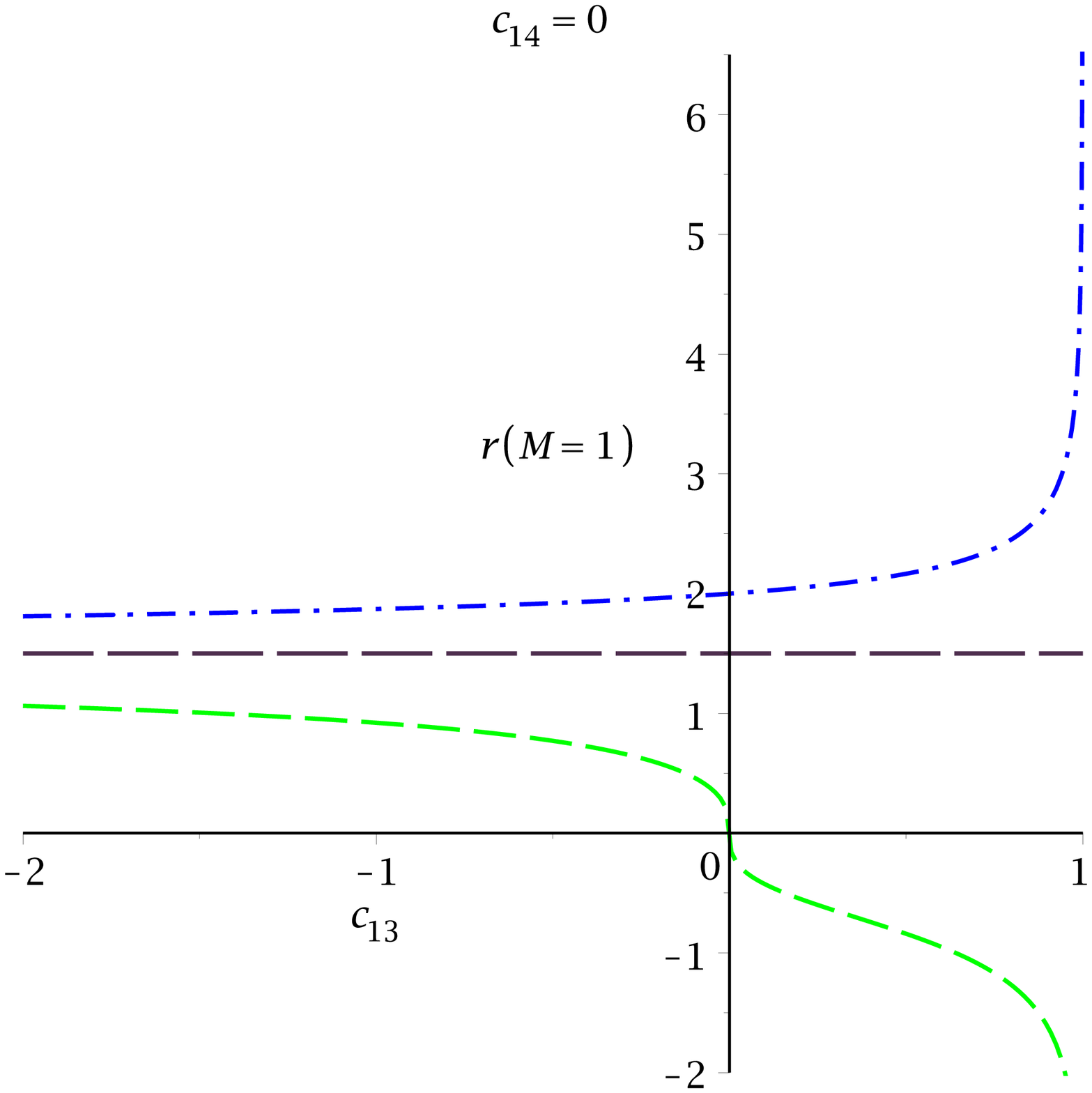}
	\includegraphics[width=7cm]{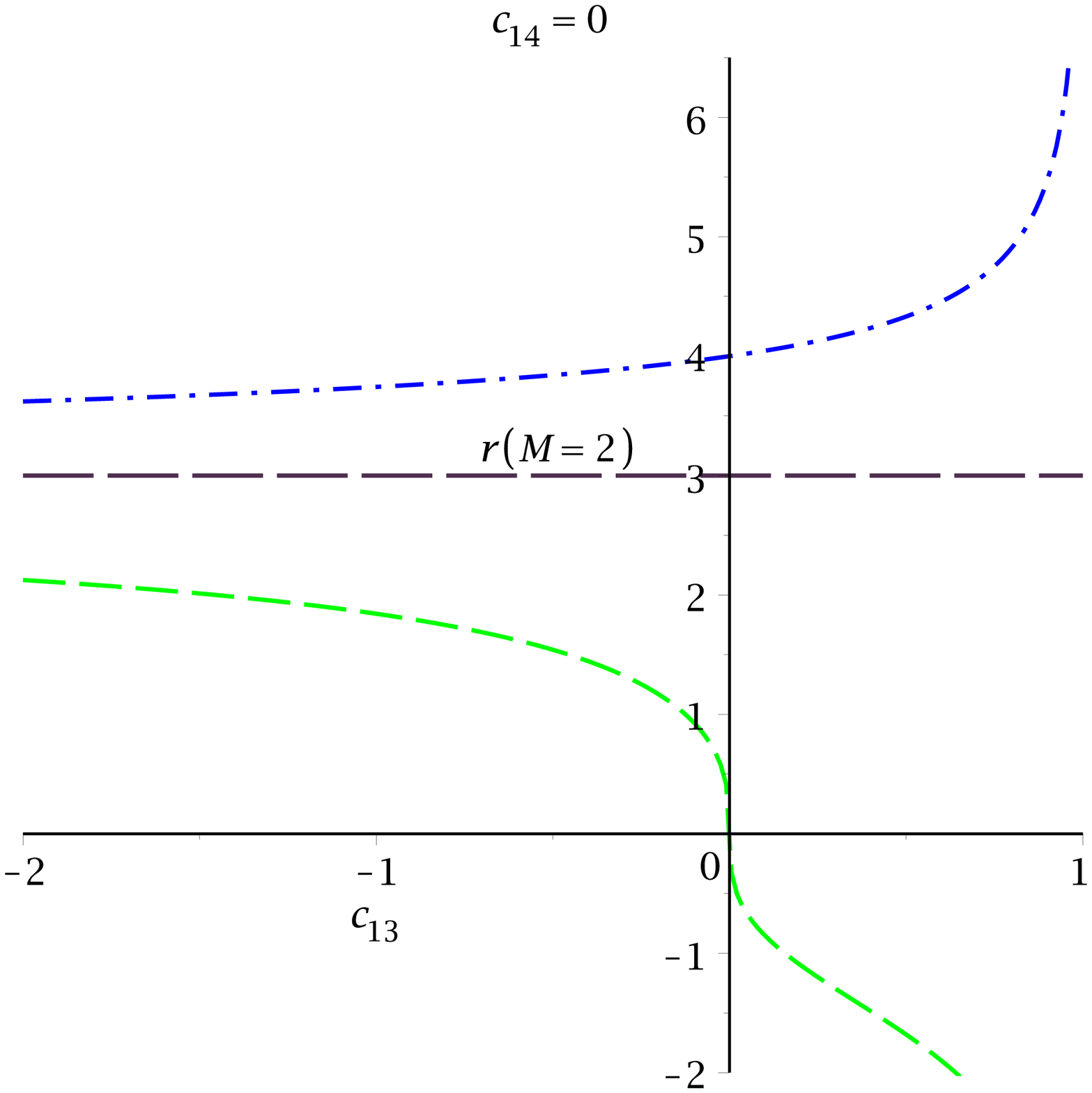}
\caption{These figures show the Killing and universal radii
for the Case \textcolor{black}{B} where we have:
	$r_{kh1}$ (blue dot-dashed line), 
	$r_{kh2}$ (green dashed line), $r_{uh1}$ (violet long-dashed). The Killing and
	universal horizons
	that are not displayed in these figures are imaginaries.}
	\label{figD1}	
\end{figure}

\textcolor{black}{
Thus, using equation (\ref{G4a}), the aether vector can be written as
\bqn
a&=&\frac{-3\sqrt{3} \zeta {M}^{2}}{4 r^2 \sqrt {1 -{c_{13}}}}
\\
b&=& \frac{ 4\epsilon \left( {c_{13}}-1 \right) \left( -2r+3M \right) {r}^{2} \sqrt{ 3{M}^{2}+4Mr+4{r}^{2}}} {\left( {c_{13}}-1 \right) 16{r}^{3} \left(  r - 2M \right)+27{c_{13}}{M}^{4}}
\eqn
}
Again, using equation (\ref{G4a}), the universal horizons are given by
\bqn
r_{uh1,2}&=&\frac{3M}{2},\\
r_{uh3,4}&=&-\frac{M}{2}+\eta_1 \frac{\sqrt {2}}{2}\sqrt {-{M}^{2}}.
\eqn
Substituting the equation (\ref{G4a}) into the Killing horizons,
we get that they depend on $c_{13}$ and the mass $M$. Thus, we plot the real 
Killing and universal horizons shown in the Figure \ref{figD1}, for two different
values of $M=1$ and $M=2$.

Since the outermost universal horizon is $r_{uh1}=r_{uh2}$, the surface gravity, temperature, entropy and the first law and
using equations (\ref{termu}) we have
\bqn
\kappa_{uh1} &=&-\frac{\epsilon\zeta \sqrt {6} }{9M} {\frac { \left|  \left( {c_{13}}+3 {c_{2}}
 \right) {c_{13}} \right| }{ \left( {c_{13}}+3 {c_{2}} \right) {
c_{13}} }}\frac{1}{\sqrt {1-{c_{13}}}}, \\
T_{uh1} &=& -\frac{\epsilon\zeta \sqrt {6}}{18\pi M} {\frac {  \left|  \left( {c_{13}}+3 {c_{2}
} \right) {c_{13}} \right| }{ \left( {c_{13}}+3 {c_{2}} \right) {
c_{13}}  }}\frac{1}{\sqrt {1-{c_{13}}}},\\
S_{uh1} &=& \frac{ 9\pi M^2 }{ 4G},\\
\delta S_{uh1} &=& -\frac{3\sqrt {6}\pi}{ \epsilon\zeta}\,\delta M\, M  {\frac {\left( {c_{13}}+3 {c_{2}} \right) 
{c_{13}}  }{ \left|  \left( {c_{13}
}+3 {c_{2}} \right) {c_{13}} \right| }}\sqrt {1-{c_{13}}}.
\eqn
Since the surface gravity must be positive, we have to choose
$-\epsilon\zeta/c_{13}(c_{13}+3 c_{2})>0$.

\section{{\bf Solutions for case \textcolor{black}{(B)}:} $\bf c_{2} \ne 0$ and $\bf c_{13} = 0$ and $\bf c_{14} = 0$}
%
For the field equations (\ref{Gtt})-(\ref{Gphiphi}) for $c_{13}=0$ and
$c_{14}=0$ are given by
\bqn
&&A=D_2+D_3 r^2+\frac{D_4}{r},\nb\\
&&B=\frac{r}{r+D_4},\nb\\
\eqn
where the two solutions for $a(r)$ and $b(r)$ can be written as
\bqn
&&a = \frac{D_1}{r^2},\nb\\
&&b =\epsilon \sqrt{\frac{D_1^2+r^4+r^3 D_4}{r^2(r+D_4)^2}},
\eqn
where $D_1$, $D_2$, $D_3$ and $D_4$ are arbitrary integration constants and
we have chosen $D_2=1$ and $D_3=0$ in order to have
a flat spacetime at infinity. The choice $D_4=-|D_4|$ is in order to have a resemblance
with the Schwarzschild solution as in the GR and $|D_4|=2M$, where $M$ is the 
Schwarzschild mass. Thus, $A(r)$, $B(r)$ and $b(r)$ can be rewritten as
\bqn
&&A=1-\frac{2M}{r},\nb\\
&&B=\frac{r}{r-2M},\nb\\
&&b =\epsilon \sqrt{\frac{D_1^2+r^4-2 r^3 M}{r^2(r-2 M)^2}},
\lb{Ab1}
\eqn
where $D_1^2+r^4-2 r^3 M > 0$. Note that the solution presented in this section does not depend explicitly on the parameter $c_{2}$, although $c_{2} \neq 0$.

The Kretschmann scalar is given by
\bq
K = \frac{ 48 M^2}{r^6}.
\eq
We can note that $r=0$ is the only singularity of this spacetime.

The Killing horizon equation is obtained from
\bqn
&&{\chi}^{\alpha} {\chi}_{\alpha} = -1+\frac{2M}{r}=0,\nb\\
\eqn
whose solution is
\bqn
&&r_{kh} = 2M.
\eqn
Note that the Killing horizon coincides just as in GR.

The universal horizon equation is
\bqn
&&{\chi}^{\alpha} {u}_{\alpha} =\epsilon \sqrt{\left(1-\frac{2M}{r}\right) \left[\frac{D_1^2}{r^3 (r-2M)}+1\right]}=0,\nb\\
\lb{uhA}
\eqn
\textcolor{black}{
whose four solutions are
\bqn
r_{uh1,2,3,4} &=& \frac{1}{2} M+\eta_1 \frac{\sqrt{3}}{12}  \alpha_2+ \nb\\
&&\eta_2 \frac{\sqrt{6}}{12}  \sqrt{\frac{12 M^2 \alpha_1 \alpha_2-\alpha_2 \alpha_1^2-48 \alpha_2 D_1^2+24 \sqrt{3} M^3  \alpha_1}{\alpha_1 \alpha_2}},
\eqn
}
where $\eta_1=\eta_2=\pm 1$  and
\bqn
&&\alpha_1=\left(432 D_1^2 M^2+12 \sqrt{-768 D_1^6+1296 D_1^4 M^4}\right)^\frac{1}{3},\nb\\
&&\alpha_2=\sqrt{\frac{12 M^2 \alpha_1+2 \alpha_1^2+96 D_1^2}{\alpha_1}}.
\eqn

Since $\alpha_1$ must be real we have that $D_1\le \sqrt{1296/768}M^2$. 
Besides, we must have that $D_1^2+r^4-2 r^3 M \ge 0$, in order to $b$ be real, as pointed out after (\ref{Ab1}). These two conditions simultaneously applied impose that 
\bq
D_1 = \sqrt{\frac{27}{16}}M^2.
\eq

\textcolor{black}{
Thus, the aether vector is given by
\bqn
a&=&{\frac {3 \sqrt {3}{M}^{2}}{4{r}^{2}}}
\\
b&=&{\frac{ \epsilon \left| -2r+3M \right| \sqrt{3{M}^{2}+4Mr+4{r}^{2}} }{4{r} \left| -r+2M \right|}}
\eqn
}

\textcolor{black}{
The Killing and universal horizons are given by
\bqn
r_{uh1,2}&=&\frac{3M}{2},\\
r_{uh3,4}&=&-\frac{M}{2}+\eta_1 \frac{\sqrt {2}}{2}\sqrt {-{M}^{2}}.
\eqn
The only real universal horizons are $r_{uh1}$ and $r_{uh2}$. Since the outermost horizon is $r_{uh1}=3M/2$, we get
for the surface gravity, temperature, entropy and the first law, using equations (\ref{termu}), thus
\bqn
\kappa_{uh1} &=& {\frac {\sqrt {6}\epsilon}{9M}}, \\
T_{uh1} &=& {\frac {\sqrt {6}\epsilon}{18M\pi }},\\
S_{uh1} &=& {\frac {9\pi \,{M}^{2}}{4G}},\\
\delta S_{uh1} &=& \frac{3\sqrt {6}\,\delta M\,M\pi}{\epsilon} .
\eqn
Since the surface gravity must be positive, we have to choose $\epsilon=1$. 
}

\section{{\bf Solutions for case \textcolor{black}{(C)}:} $\bf c_{2} = 0$  and $\bf c_{13} \neq 0$ and $\bf c_{14} = 0$ }
%
The solution of the field equations (\ref{Gtt})-(\ref{Gphiphi}) for $c_{2}=0$ and
$c_{14}=0$ we get, 
\bqn
&&A=E_1+\frac{E_2}{r^4}+E_3 r^2+\frac{E_4}{r},\nb\\
&&B=\frac{1}{1+\frac{E_2}{r^4}+\frac{E_4}{r}},\nb\\
\eqn
where the solutions of $a(r)$ and $b(r)$ are given by
\bqn
&&a = \zeta \sqrt{-\frac{E_2}{c_{13}\;r^4}},\nb\\
&&b = \epsilon \sqrt{\frac{r^4 \left(-E_2+c_{13} r^4+c_{13} E_2-2 c_{13} M r^3\right)}{c_{13} \left(r^4+E_2-2 M r^3\right)^2}},
\eqn
where $\zeta=\pm 1$, $E_1$, $E_2$, $E_3$ and $E_4$ are arbitrary integration constants 
we have chosen $E_1=1$ and $E_3=0$ in order to have
a flat spacetime at infinity and $E_4=-|E_4|$ in order to have a resemblance
with the Schwarzschild solution as in the GR and $|E_4|=2M$, where $M$ is the 
Schwarzschild mass. Thus, $A(r)$, $B(r)$ and $b(r)$ can be rewritten as
\bqn
&&A=1+\frac{E_2}{r^4}-\frac{2M}{r},\nb\\
&&B=\frac{1}{1+\frac{E_2}{r^4}-\frac{2M}{r}},\nb\\
&&b = \epsilon \sqrt{\frac{r^4 \left(-E_2+c_{13} r^4+c_{13} E_2-2 c_{13} M r^3\right)}{c_{13} \left(r^4+E_2-2M r^3\right)^2}},
\eqn
where $-{E_2}/{c_{13}} > 0$ and $\left(-E_2+c_{13} r^4+c_{13} E_2-2 c_{13} M r^3\right)/c_{13} > 0$.
The first condition is imposed in order to have $a$ real, while the second one ensures that $b$ be real. The last one is not obvious but it can be found if we analyze the first and second derivatives of the expression inside the parenthesis, beyond it behavior in the limits for $r\rightarrow \pm \infty$ and the position of the minimum of the function.

Note that the solutions presented in this section depend explicitly on the parameter $c_{13}$. Besides,
the term $r^{-4}$ is interesting because it could be compared in GR  to the 
Hartle and Thorne \cite{Hartle1968} solution of a slowly rotating deformed relativistic star,
assuming the quadrupole moment is zero.

The Kretschmann scalar is given by
\bq
K = \frac{12 (39 E_2^2-20 E_2 M r^3+4 M^2 r^6)}{r^{12}}.
\eq
Notice again that $r=0$ is the only singularity.

The Killing horizon equation is given by
\bqn
&&{\chi}^{\alpha} {\chi}_{\alpha} = -1-\frac{E_2}{r^4}+\frac{2M}{r}=0,\nb\\
\eqn
whose four roots are
\bqn
&&r_{kh1,2,3,4} = \frac{1}{2} M+\eta_1\frac{\sqrt{3}}{12}  \beta_2+\nb\\
&&\eta_2 \frac{1}{12} \sqrt{-\frac{-72 M^2 \beta_1 \beta_2+6 \beta_2 \beta_1^2+
288 \beta_2 E_2-144 \sqrt{3} M^3  \beta_1}{\beta_1 \beta_2}},
\eqn
where 
\bqn
&&\beta_1=\left(432 E_2 M^2+12 \sqrt{-768 E_2^3+1296 E_2^2 M^4}\right)^\frac{1}{3},\nb\\
&&\beta_2=\sqrt{\frac{12 M^2 \beta_1+2 \beta_1^2+96 E_2}{\beta_1}}.
\eqn

The universal horizon is
\bqn
&&{\chi}^{\alpha} {u}_{\alpha} =\epsilon \sqrt{-\frac{E_2-c_{13} r^4-c_{13} E_2+2 c_{13} M r^3}{c_{13} r^4}}=0,\nb\\
\lb{uhB}
\eqn
whose four solutions are
\bqn
&&r_{uh1,2,3,4} = \frac{1}{2} M+\eta_1 \frac{\sqrt{3}}{12}  \gamma_3+ \nb\\
&& \eta_2 \frac{\sqrt{6}}{12\sqrt{\gamma_2 \gamma_3}}  \left[ 12 M^2 c_{13}\gamma_2 \gamma_3-12^\frac{1}{3}\gamma_3  \gamma_2^2+4\;.\;12^\frac{2}{3} \gamma_3 E_2  c_{13}-4\;.\;12^\frac{2}{3} \gamma_3 E_2  c_{13}^2+\right.\nb\\
&&\left. 24 \sqrt{3} M^3  c_{13} \gamma_2 \right]^\frac{1}{2},
\eqn
where 
\bqn
&&\gamma_1= 36 M^2+\sqrt{3} \sqrt{\frac{256 E_2-256 c_{13} E_2+432 M^4 c_{13}}{c_{13}}},\nb\\
&&\gamma_2=\left[ (-1+c_{13})E_2 \gamma_1 c_{13}^2\right]^\frac{1}{3},\nb\\
&&\gamma_3=\sqrt{\frac{12 M^2 c_{13} \gamma_2+2\;.\; 12^\frac{1}{3} \gamma_2^2-8\;.\;12^\frac{2}{3} E_2  c_{13}+8\;.\;12^\frac{2}{3} E_2  c_{13}^2}{c_{13} \gamma_2}}.
\eqn

In order to have real Killing horizons we must impose that
\bq
E_2 \le  \frac{27}{16} M^4,
\eq
and real universal horizons we must have
\bq 
E_2 \ge  \frac{27}{16} \left( \frac{c_{13}}{c_{13}-1} \right) M^4.
\eq
From these two conditions we have that
\bq 
E_2 =  \frac{27}{16} \left( \frac{c_{13}}{c_{13}-1} \right) M^4.
\lb{E2}
\eq

Thus, using equation (\ref{E2}), the aether vector can be written as
\bqn
a&=&\frac{3\sqrt{3} \zeta {M}^{2}}{4 r^2 \sqrt {1 -{c_{13}}}}
\\
b&=& \frac{ 4\epsilon \left( {c_{13}}-1 \right) \left( -2r+3M \right) {r}^{2} \sqrt{ 3{M}^{2}+4Mr+4{r}^{2}}} {\left( {c_{13}}-1 \right) 16{r}^{3} \left(  r - 2m \right)+27{c_{13}}{M}^{4}}
\eqn

\begin{figure}[!ht]
\centering
	\includegraphics[width=7cm]{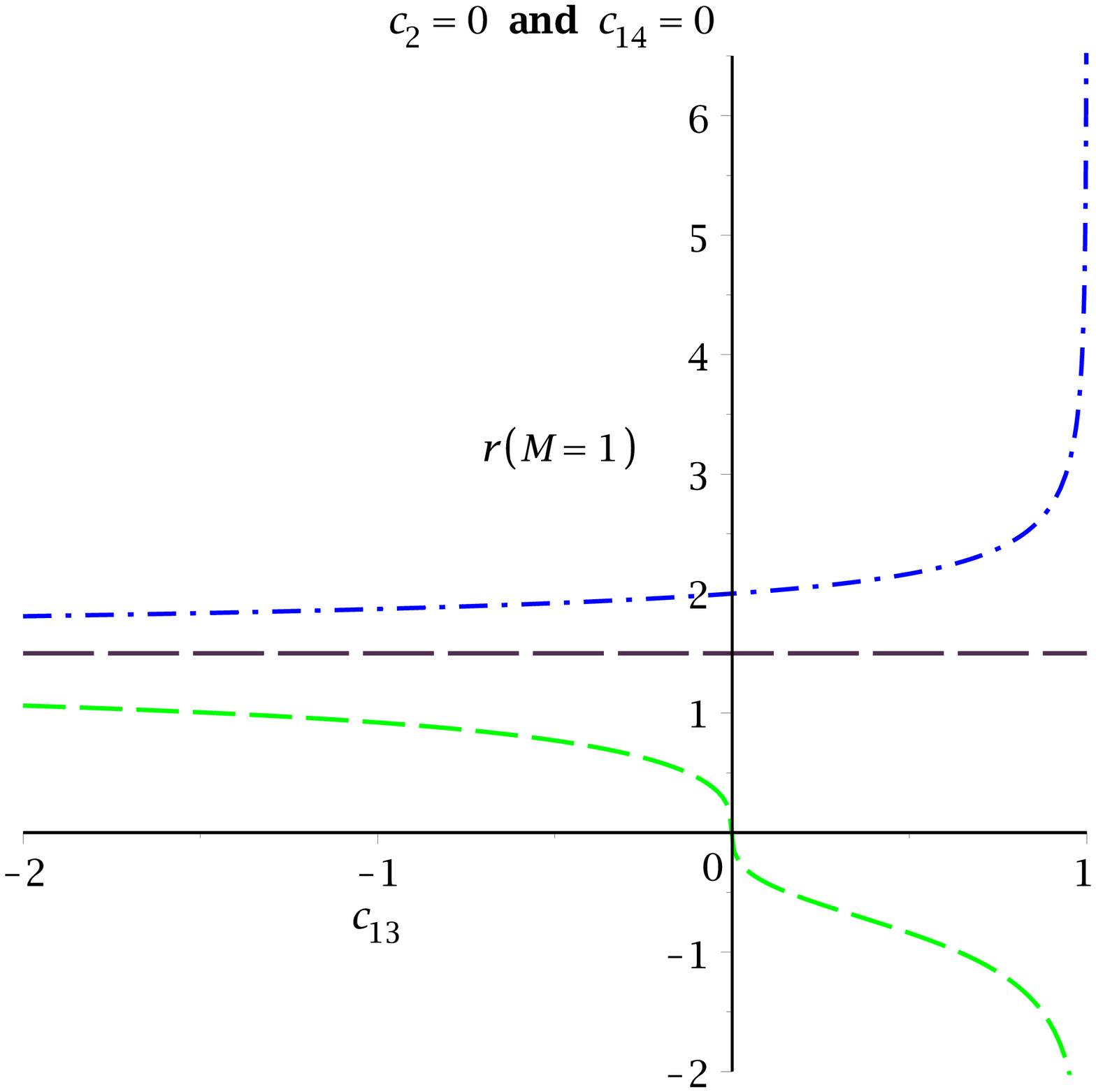}
	\includegraphics[width=7cm]{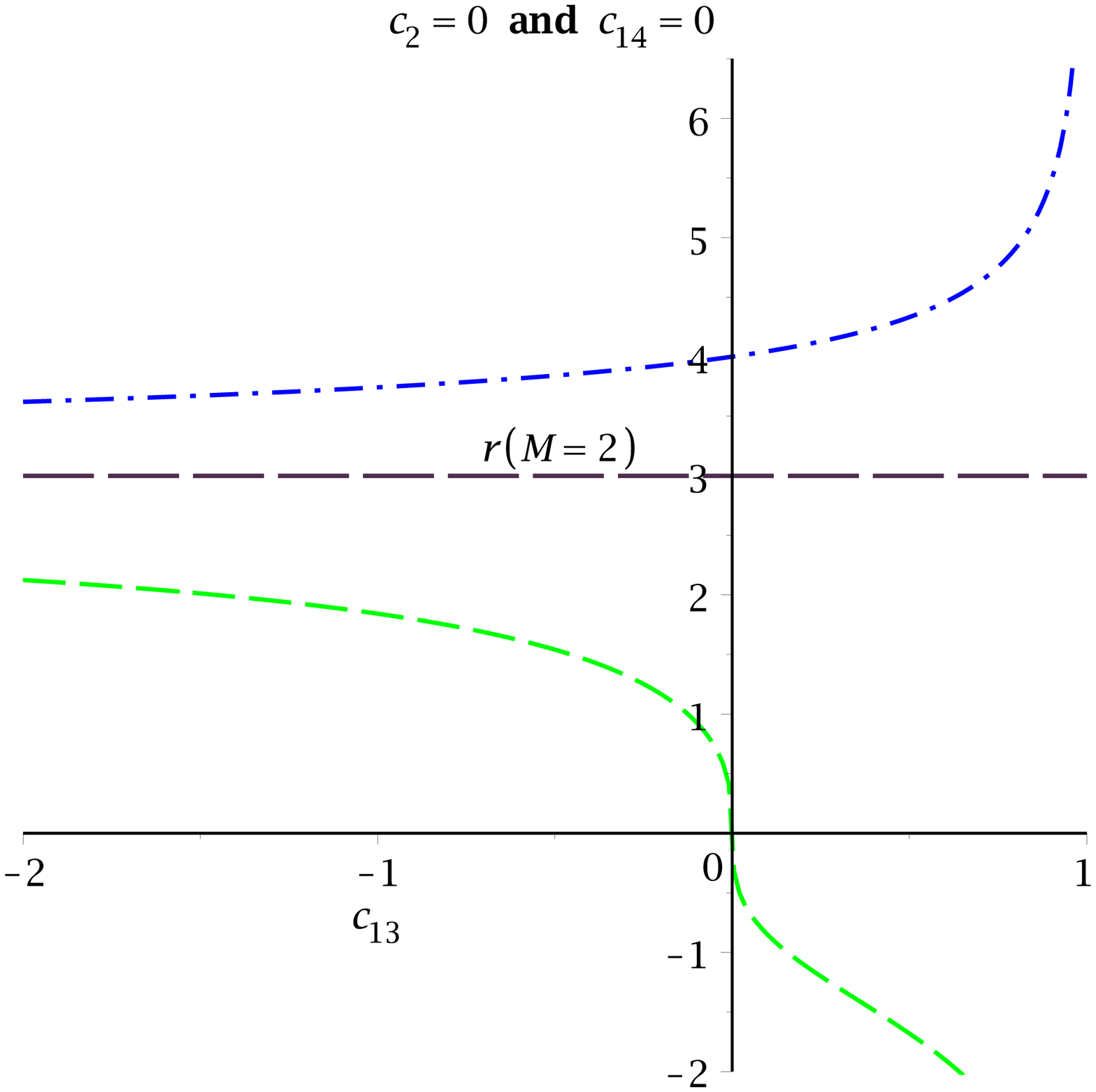}
\caption{These figures show the Killing and universal radii
for the Case \textcolor{black}{C}, where we have:
	$r_{kh1}$ (blue dot-dashed line), 
	$r_{kh2}$ (green dashed line), $r_{uh1}$ (violet long-dashed). The Killing and
	universal horizons
	that are not displayed in these figures are imaginaries.}
	\label{figB}	
\end{figure}

Again, using equation (\ref{E2}), the universal horizons are given by
\bqn
r_{uh1,2}&=&\frac{3M}{2},\\
r_{uh3,4}&=&-\frac{M}{2}+\eta_1 \frac{\sqrt {2}}{2}\sqrt {-{M}^{2}}.
\eqn
Substituting the equation (\ref{E2}) into the Killing horizons,
we get that they depend on $c_{13}$ and the mass $M$. Thus, we plot the real 
Killing and universal horizons shown in the Figure \ref{figB}, for two different
values of $M=1$ and $M=2$.

Since from Figure \ref{figB} the outermost universal horizon is $r_{uh1}$, the surface gravity, temperature, entropy and the first law and
using equations (\ref{termu}) we have
\textcolor{black}{
\bqn
\kappa_{uh1} &=& {\frac {\sqrt {6}}{9M\sqrt {1-{c_{13}}}}},\\
T_{uh1} &=& {\frac {\sqrt {6}}{18\pi M\sqrt {1-{c_{13}}} }},\\
S_{uh1} &=& {\frac {9\pi \,{M}^{2}}{4G}},\\
\delta S_{uh1} &=& 3\,\sqrt {6}\,\pi \delta M\,M\sqrt {1-{c_{13}}}.
\eqn
Since the surface gravity must be positive and real, we have to choose
$\epsilon=\eta=1$ and $c_{13}<1$.
}

\section{{\bf Solutions for case \textcolor{black}{(D)}:} $\bf c_{2} = 0$ and $\bf c_{13} = 0$ and $\bf c_{14} \neq 0$ }
%
The solution of the field equations (\ref{Gtt})-(\ref{Gphiphi}) for $c_{2}=0$ and
$c_{13}=0$ we get, 
\bqn
&&A=F_1+\frac{F_2}{r}+\frac{F_3}{r^2},\nb\\
&&B=\frac{r^2}{r^2+F_2 r+F_3},\nb\\
\eqn
where $F_1$, $F_2$, $F_3$ and $F_4$ are arbitrary integration constants
we have chosen $F_1=1$ in order to have
a flat spacetime at infinity and $F_2=-|F_2|$ in order to have a resemblance
with the Schwarzschild solution as in the GR and $|F_2|=2M$, where $M$ is the 
Schwarzschild mass. Thus, $A(r)$ and $B(r)$ can be rewritten as
\bqn
&&A=1-\frac{2M}{r}+\frac{F_3}{r^2},\nb\\
&&B=\frac{r^2}{r^2-2M r+F_3}.\nb\\
\eqn
The metric of this case can be associated to the Reissner-Nordstr\"{o}m spacetime in GR, identifying $F_3$ with the electric charge.

The Kretschmann scalar is given by
\bq
K = \frac{4 (12 M^2 r^2-24 M F_3 r +14 F_3^2)}{r^8}.
\eq
We can notice again that $r=0$ is the only singularity.

The solutions for $a(r)$ and $b(r)$ are
\bqn
a&=&\frac{\zeta}{c_{14} r} 
\sqrt{c_{14}( 2Mc_{14} r-c_{14} r^2+F_3 r^2 F_4^2 c_{14}^2+2\sqrt{2} F_3  F_4 c_{14} r-F_3 c_{14}+2 F_3)},\nb\\
b &=& \frac{\epsilon\; r^2}{(r^2-2M r+F_3)} \sqrt{\frac{F_3 (r^2 F_4^2 c_{14}^2+2 \sqrt{2} F_4 c_{14} r+2)}{c_{14} r^2}},
\eqn
where $( 2M r-c_{14} r^2+F_3 r^2 F_4^2 c_{14}^2+2\sqrt{2} F_3  F_4 c_{14} r-F_3 c_{14}+2 F_3) > 0$ and $F_3 (r^2 F_4^2 c_{14}^2+2 \sqrt{2} F_4 c_{14} r+2) > 0$, in order to have $a$ and $b$ real.

Looking at the term under the square root at $b$, we can see that it vanishes only at $r=-\frac{\sqrt{2}}{F_4C_{14}}$, which coincides with the minimum or maximum of the function, depending on whether the sign of $F_4$ is positive or negative, respectively. If $F_4>0$, the minimum occurs for $r<0$, while if  $F_4<0$, the maximum occurs for $r>0$.  Thus, to ensure that $b$ is real for every value of $r \ge 0$, we must choose $F_3 > 0$ and $F_4 > 0$.
Doing a similar analysis of the term under the square root in $a$ we see that it has a minimum, that occurs for some $r<0$ or a maximum, that occurs for some $r>0$, depending on whether $F_3 > \frac{1}{{F_4}^2 c_{14}}$ or if $F_3 < \frac{1}{{F_4}^2 c_{14}}$, respectively. Imposing, as for $b$, that $a$ is real for all $r\geq 0$, the only possible option is $F_3 > \frac{1}{{F_4}^2 c_{14}}$ . So, the existence of the aether vector field imposes ${F_4}>0$ and $F_3 > \frac{1}{{F_4}^2 c_{14}}$.

We can factorize these aether component equations giving
\bqn
&&a=\frac{\epsilon}{r} \sqrt{ \frac{F_3(c_{14} F_4 r + \sqrt{2})^2-c_{14}(r^2-2M r+F_3)}{c_{14}}},\nb\\
&&b =\frac{\epsilon\; r (c_{14} F_4 r + \sqrt{2}) }{(r^2-2M r+F_3)} \sqrt{\frac{F_3}{c_{14}}}.
\eqn

Note that the solutions presented in this section depend explicitly on the parameter $c_{14}$.

The Killing horizon equation is
\bqn
&&{\chi}^{\alpha} {\chi}_{\alpha} = -1+\frac{2M}{r}-\frac{F_3}{r^2}=0,\nb\\
\eqn
whose two roots are
\bqn
&&r_{kh1,2} = M+ \eta_1 \sqrt{M^2- F_3},
\eqn
where $F_3 \le M^2$,
establishing an upper limit for the constant $F_3$, to insure that exist Killing horizons. Then, the solution for this case is restricted to values of $F_3$ and $F_4$ such that $F_4 > 0$ and 
\bq
\frac{1}{{F_4}^2 c_{14}} < F_3 \le M^2.
\lb{cond1}
\eq 

The universal horizon equation is given by
\bqn
&&{\chi}^{\alpha} {u}_{\alpha} =\epsilon \sqrt{1-\frac{2M}{r}+\frac{F_3}{r^2}}\times \nb\\ 
&&\sqrt{\frac{2 c_{14} M r-c_{14} r^2+F_3 r^2 F_4^2 c_{14}^2+2\sqrt{2} F_3  F_4 c_{14} r-F_3 c_{14}+2 F_3}{c_{14} (r^2-2M r+F_3)}+1}=0,\nb\\
\lb{uhC}
\eqn
whose three solutions are
\bqn
&&r_{uh1} = -\frac{\sqrt{2}}{F_4 c_{14}},\nb\\
&&r_{uh2,3} = M+ \eta_1 \sqrt{ M^2- F_3}.
\eqn

\begin{figure}[!ht]
\centering
	\includegraphics[width=6.5cm]{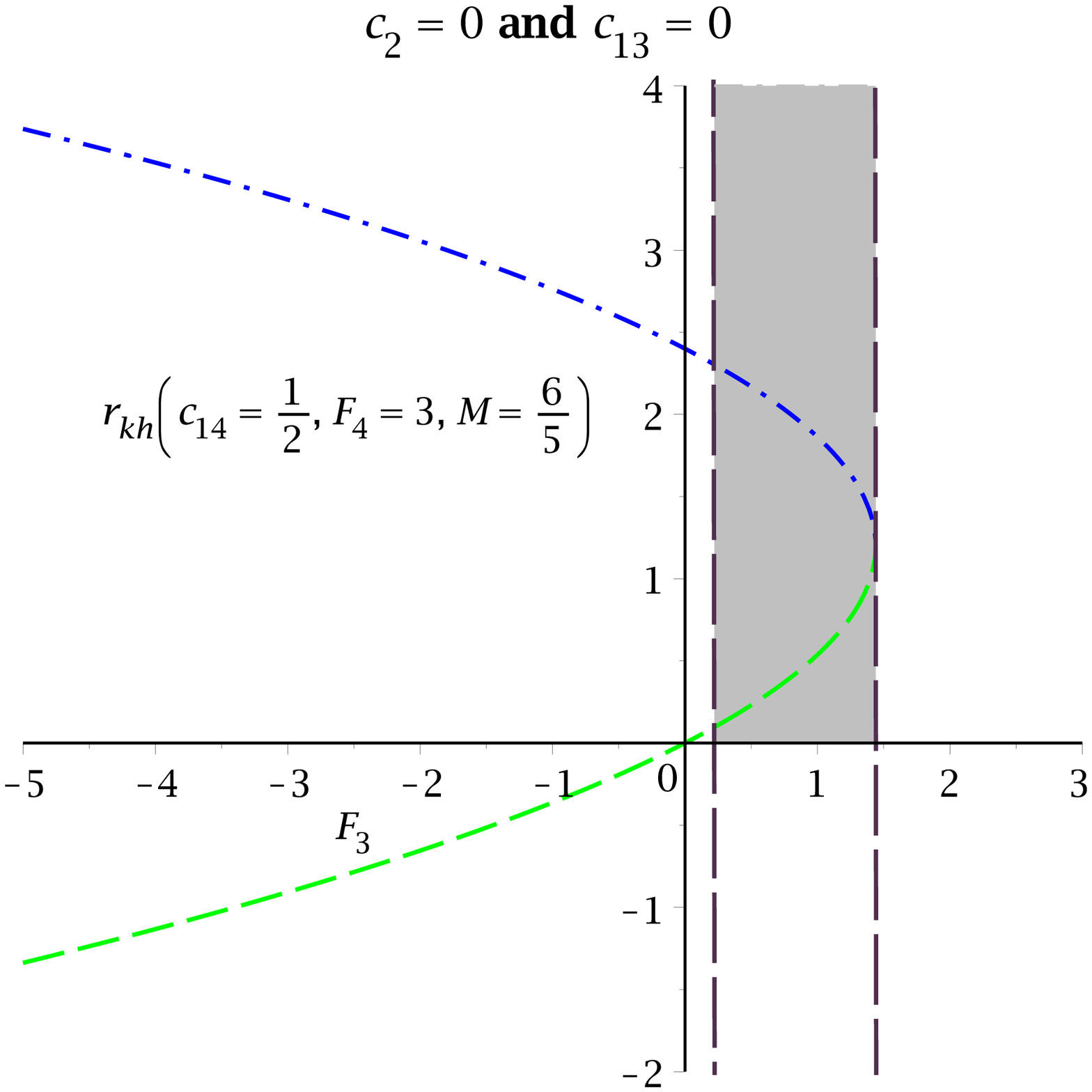}
	\includegraphics[width=6.5cm]{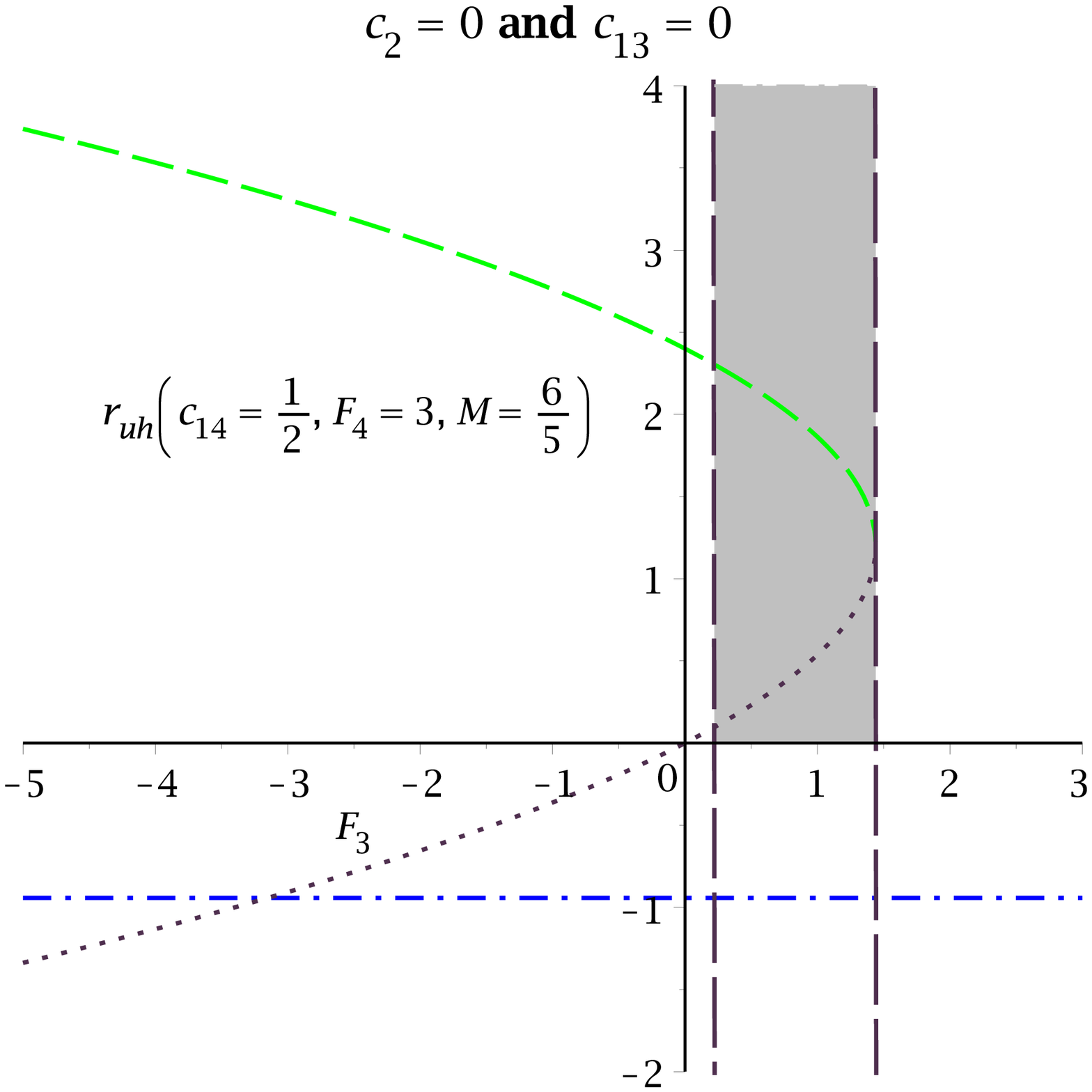}
	\includegraphics[width=6.5cm]{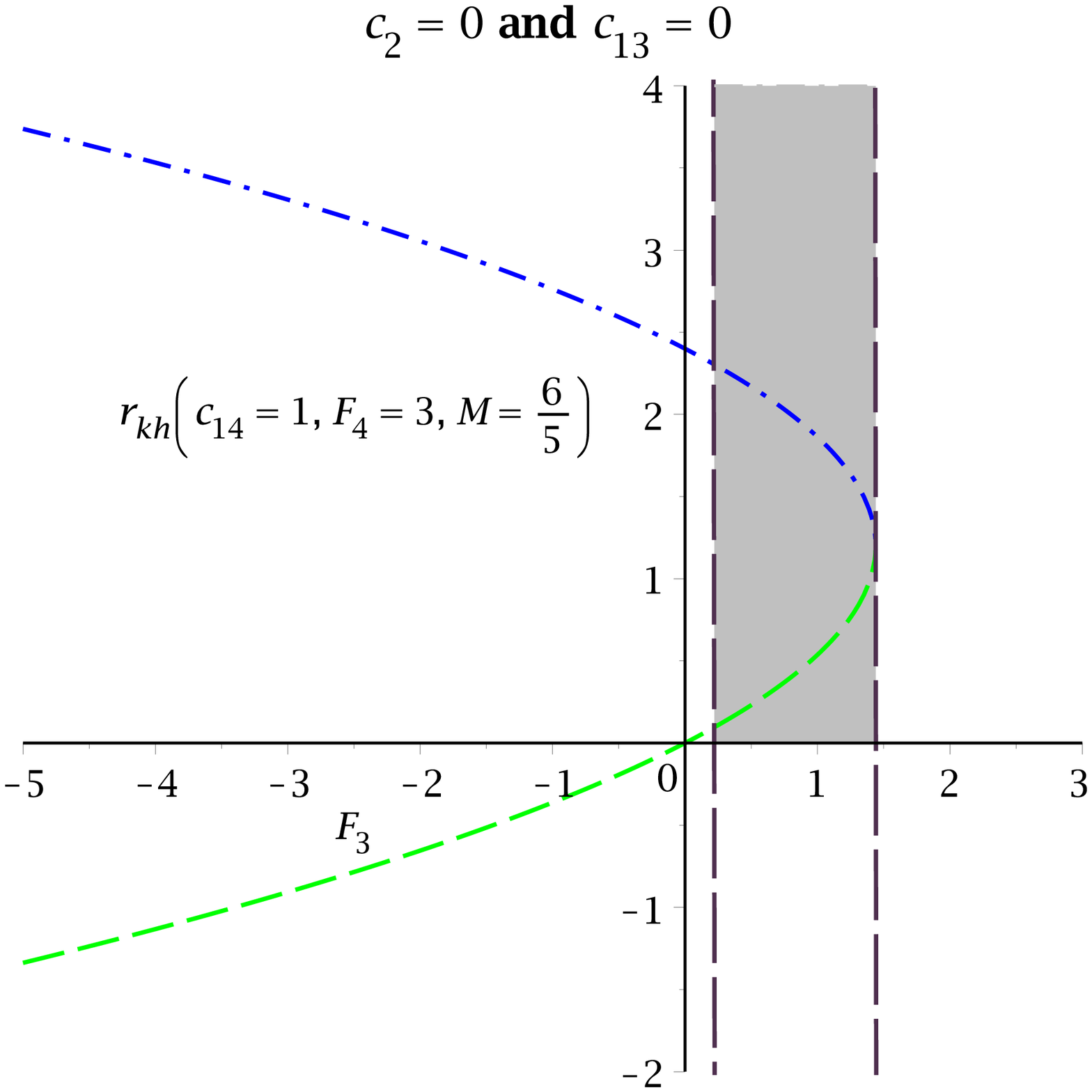}
	\includegraphics[width=6.5cm]{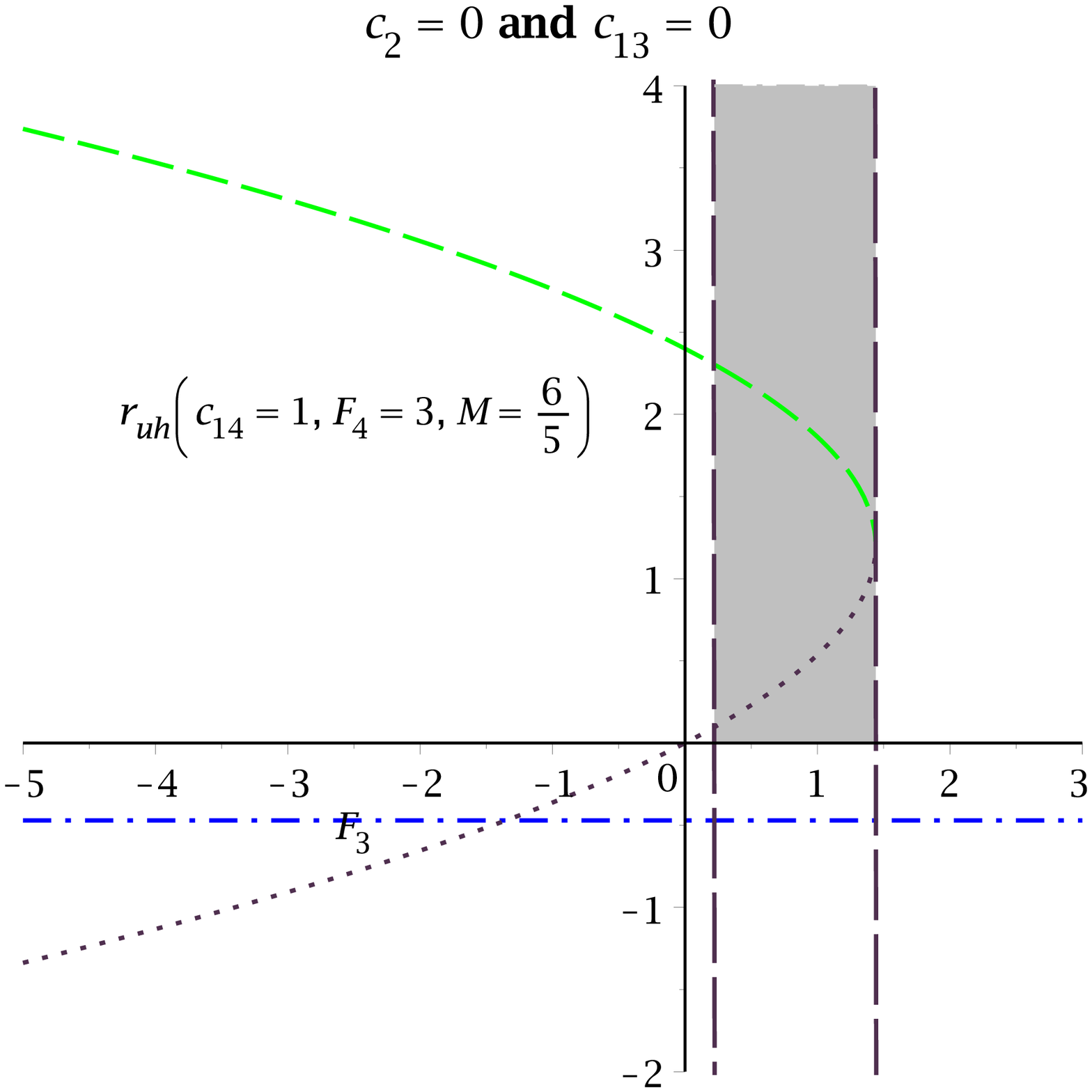}
	\includegraphics[width=6.5cm]{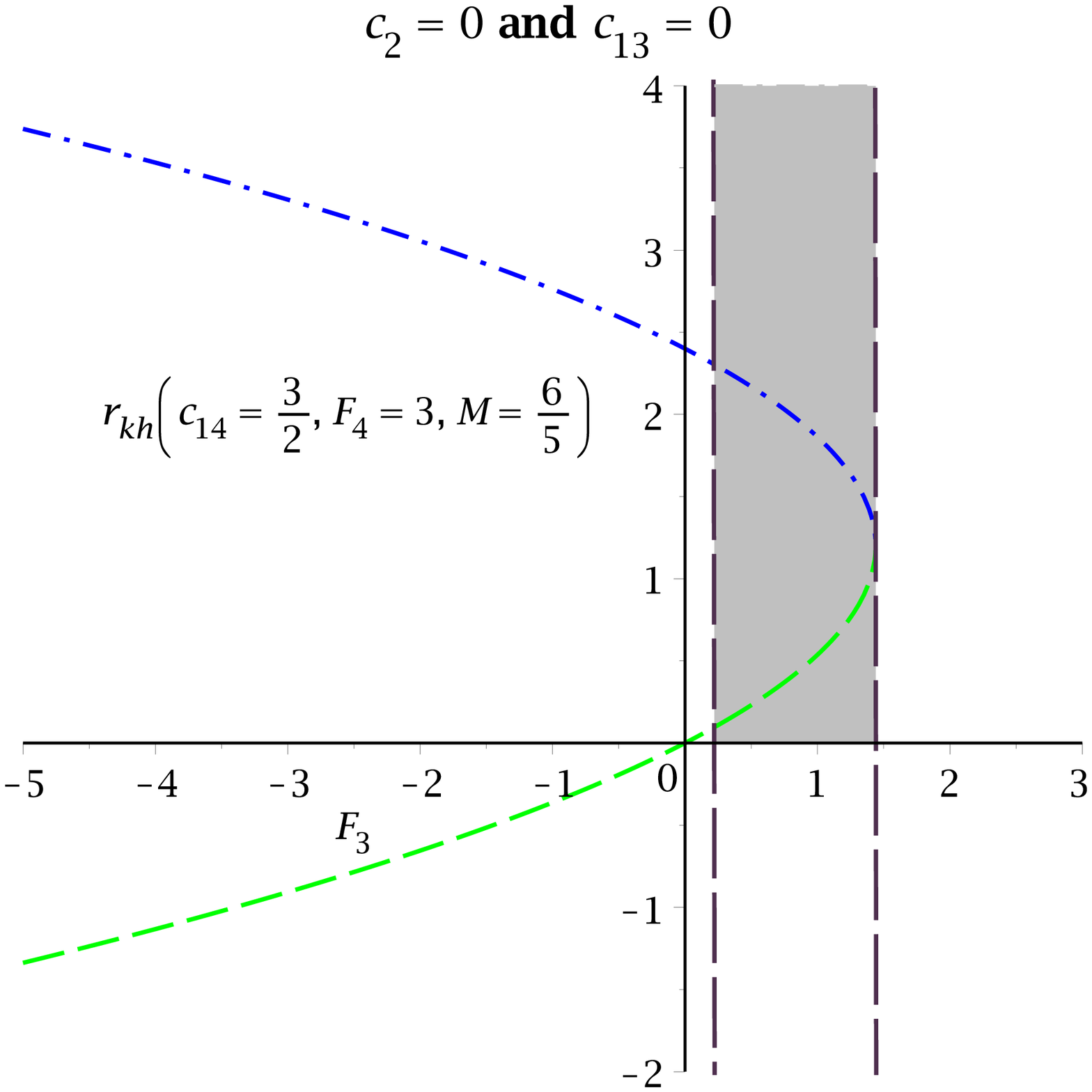}
	\includegraphics[width=6.5cm]{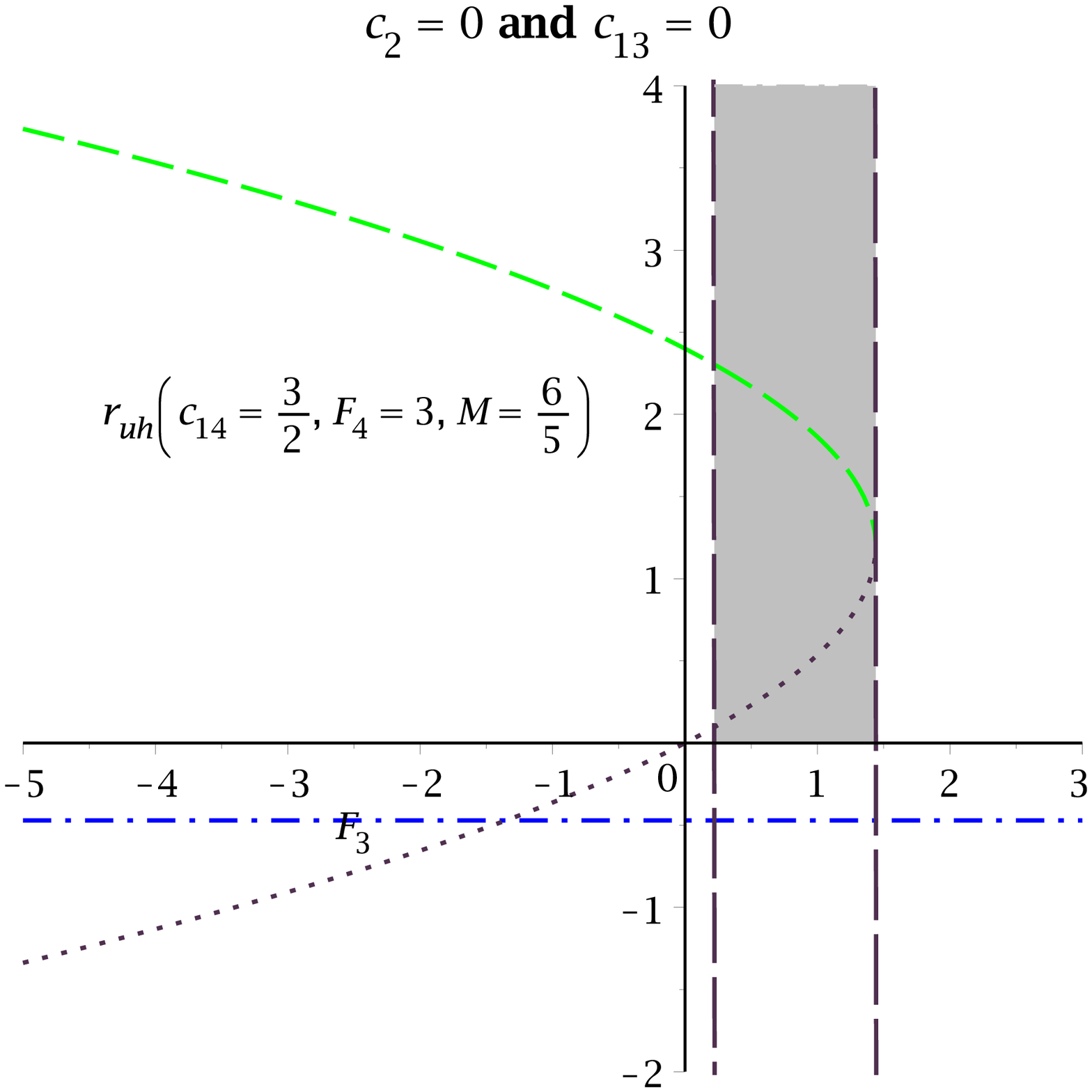}
\caption{These figures show the Killing and universal radii 
for the Case \textcolor{black}{D} where we have:
	$r_{kh1}$ or $r_{uh1}$ (blue dot-dashed line), 
	$r_{kh2}$ or $r_{uh2}$ (green dashed line), $r_{kh3}$ or $r_{uh3}$ (black dotted line). The violet long-dashed straight lines
	represent the inferior and superior limits of $F_3$. The horizons
	that are not displayed in these figures are imaginaries.
	The gray areas are the regions where the condition (\ref{cond1}) is valid.}
	\label{figC1}	
\end{figure}

\begin{figure}[!ht]
\centering
	\includegraphics[width=6.5cm]{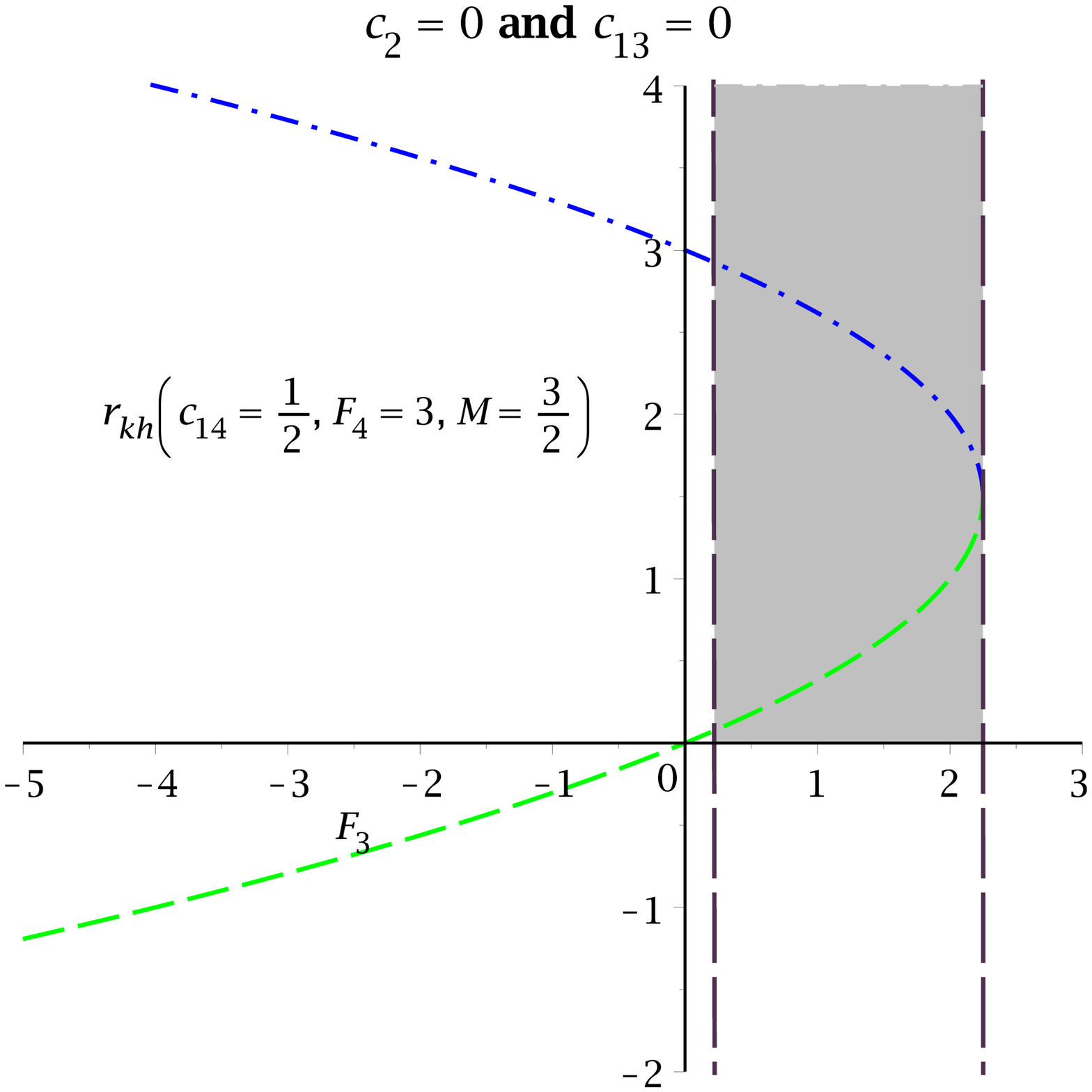}
	\includegraphics[width=6.5cm]{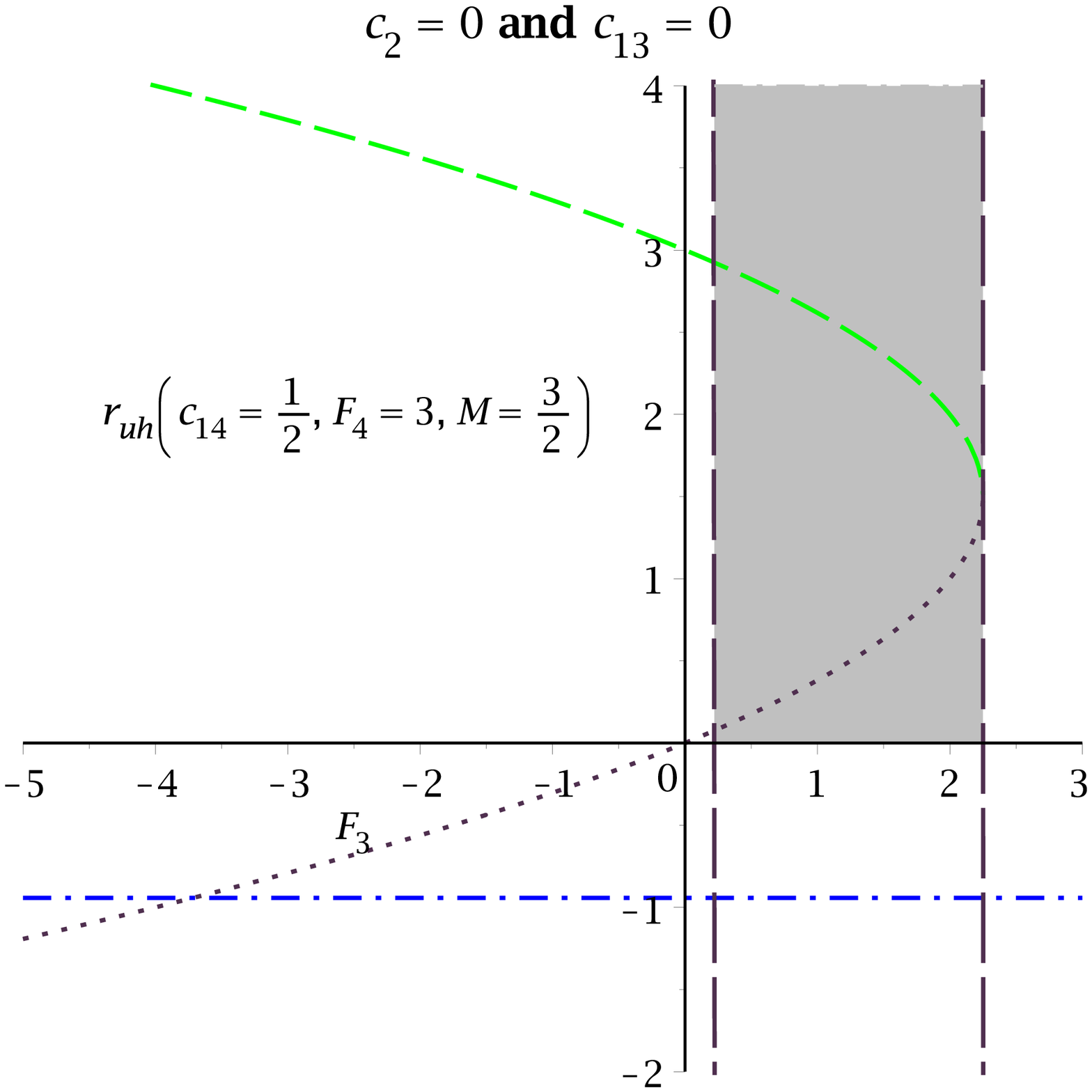}
	\includegraphics[width=6.5cm]{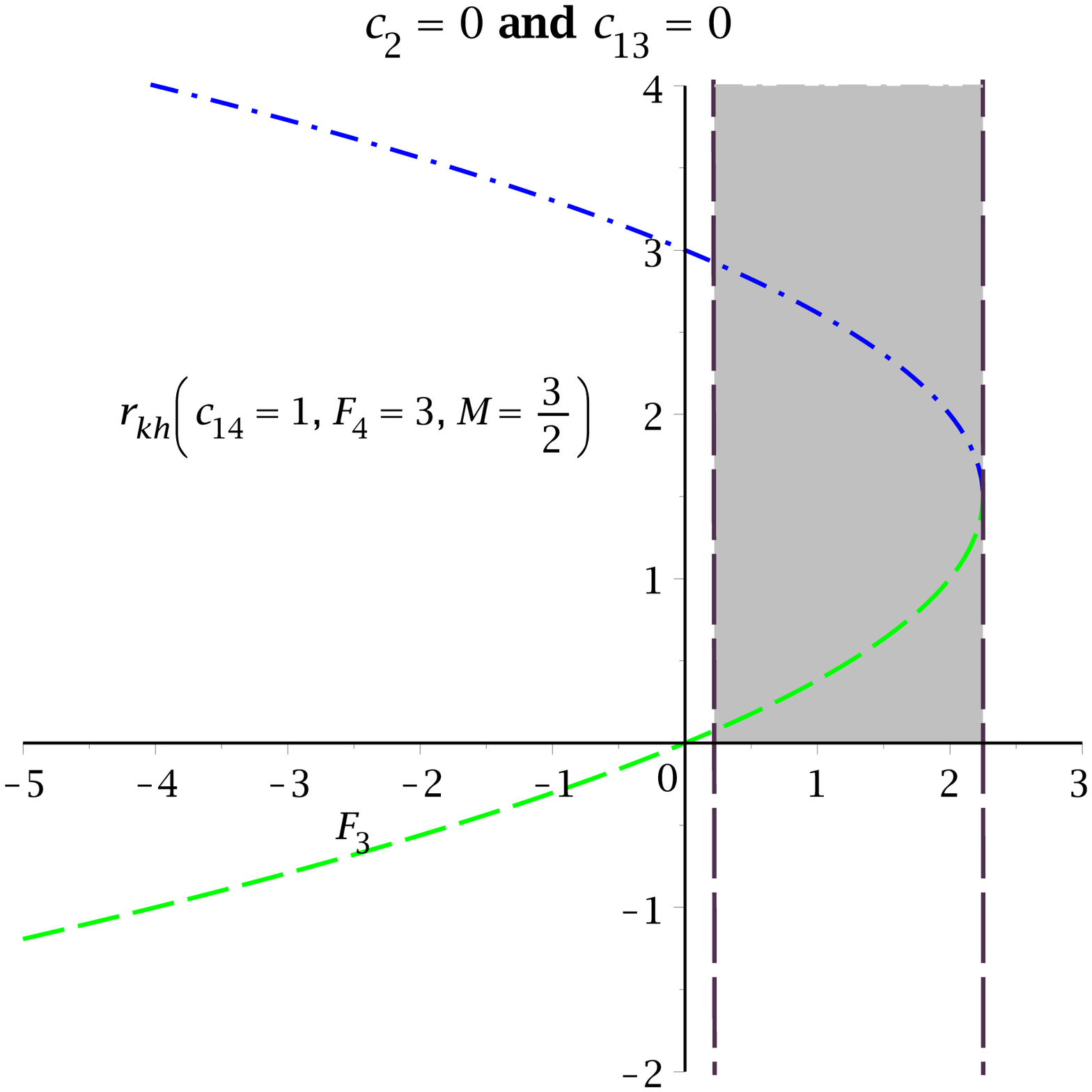}
	\includegraphics[width=6.5cm]{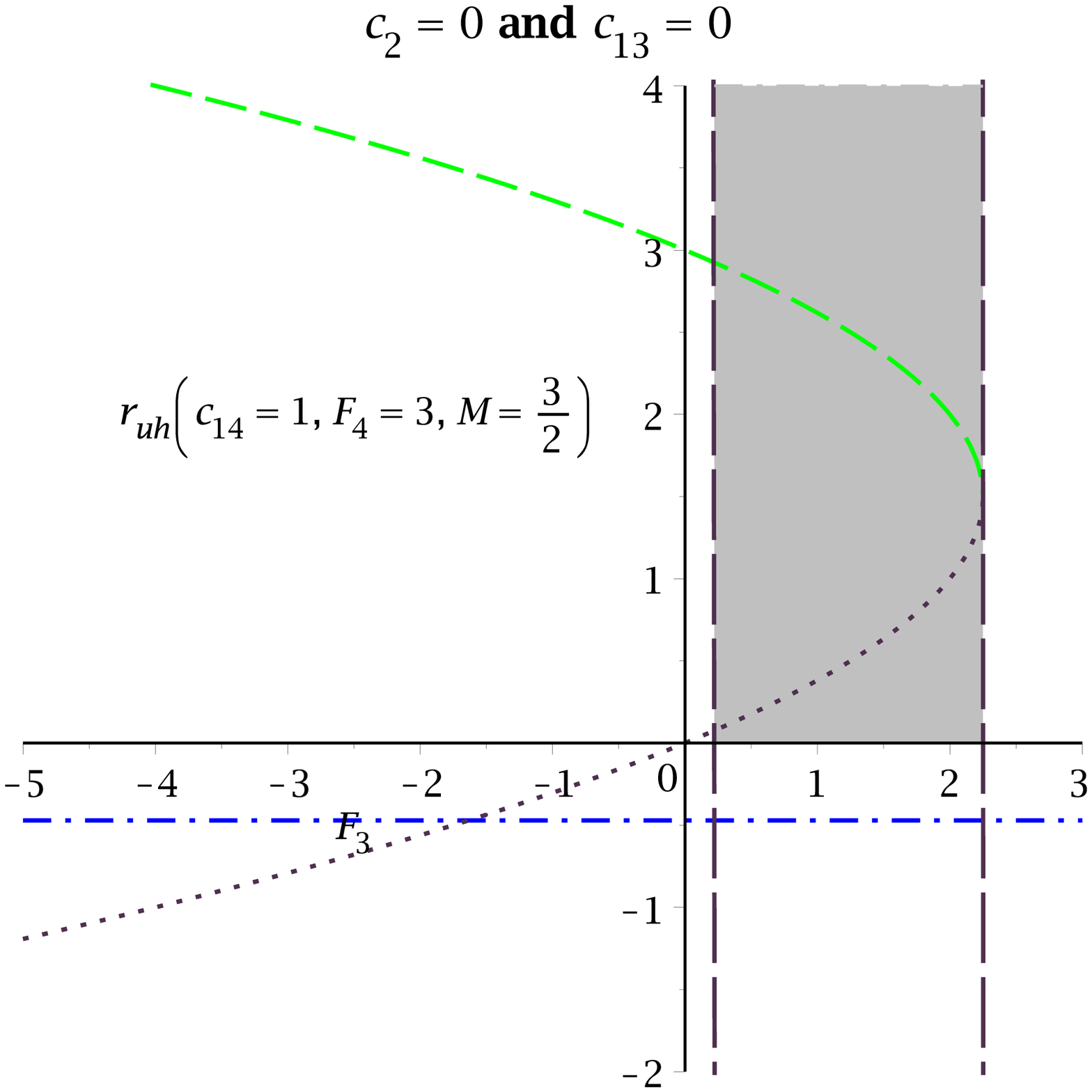}
	\includegraphics[width=6.5cm]{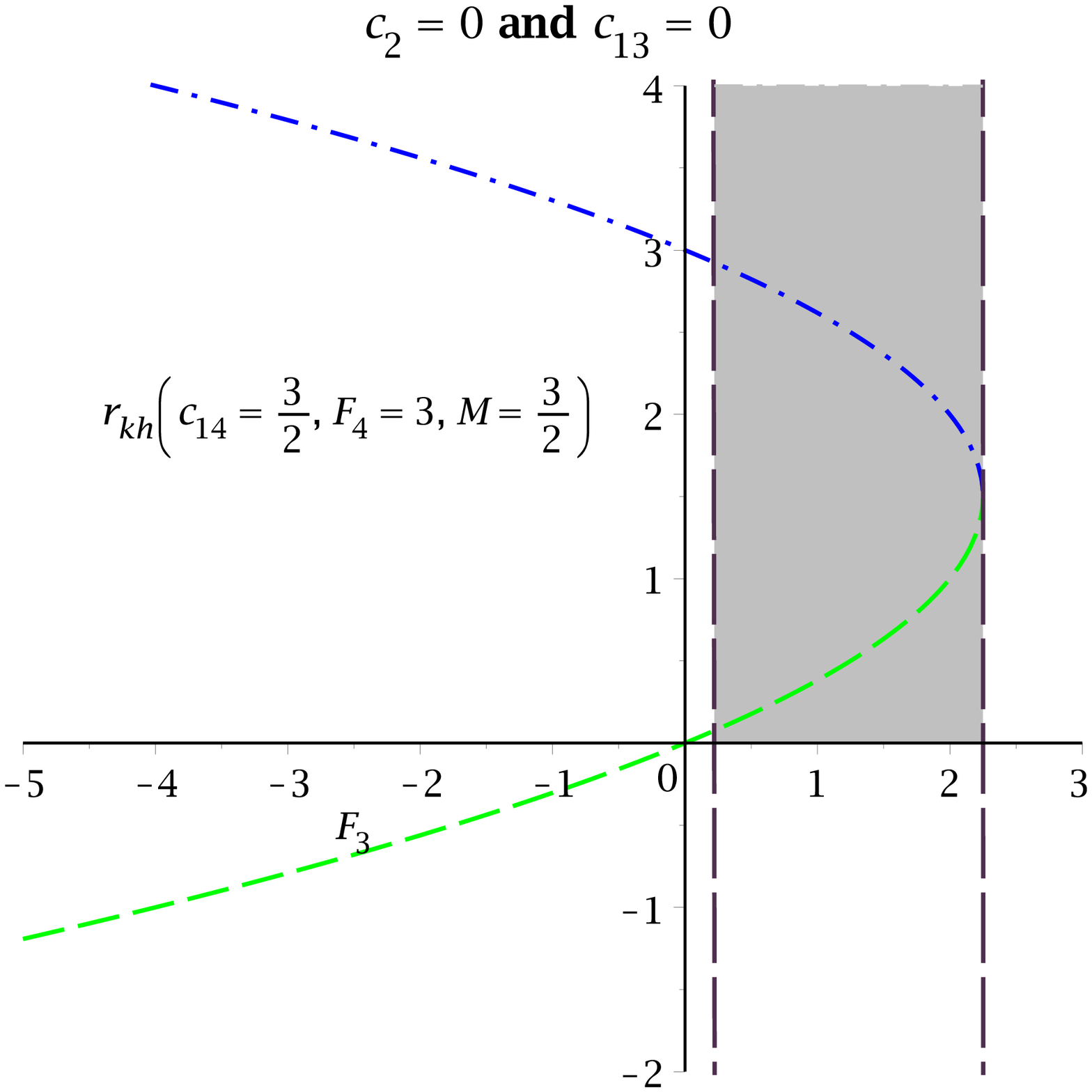}
	\includegraphics[width=6.5cm]{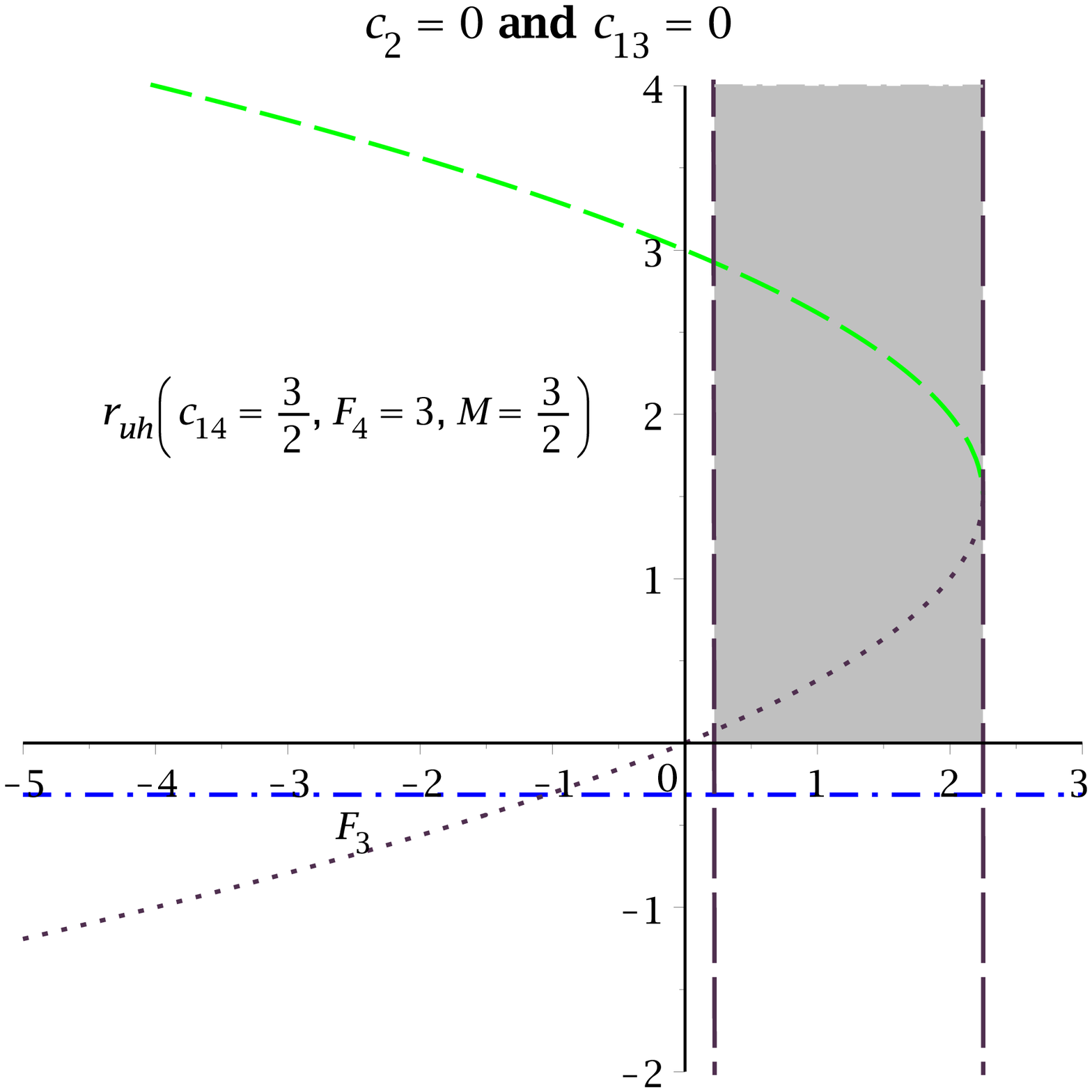}
\caption{These figures show the Killing and universal radii
for the Case \textcolor{black}{D} where we have:
	$r_{kh1}$ or $r_{uh1}$ (blue dot-dashed line), 
	$r_{kh2}$ or $r_{uh2}$ (green dashed line), $r_{kh3}$ or $r_{uh3}$ (black dotted line). The violet long-dashed straight lines
	represent the inferior and superior limits of $F_3$. The horizons
	that are not displayed in these figures are imaginaries.
	The gray areas are the regions where the condition (\ref{cond1}) is valid.}
	\label{figC2}	
\end{figure}

\begin{figure}[!ht]
\centering
	\includegraphics[width=6.5cm]{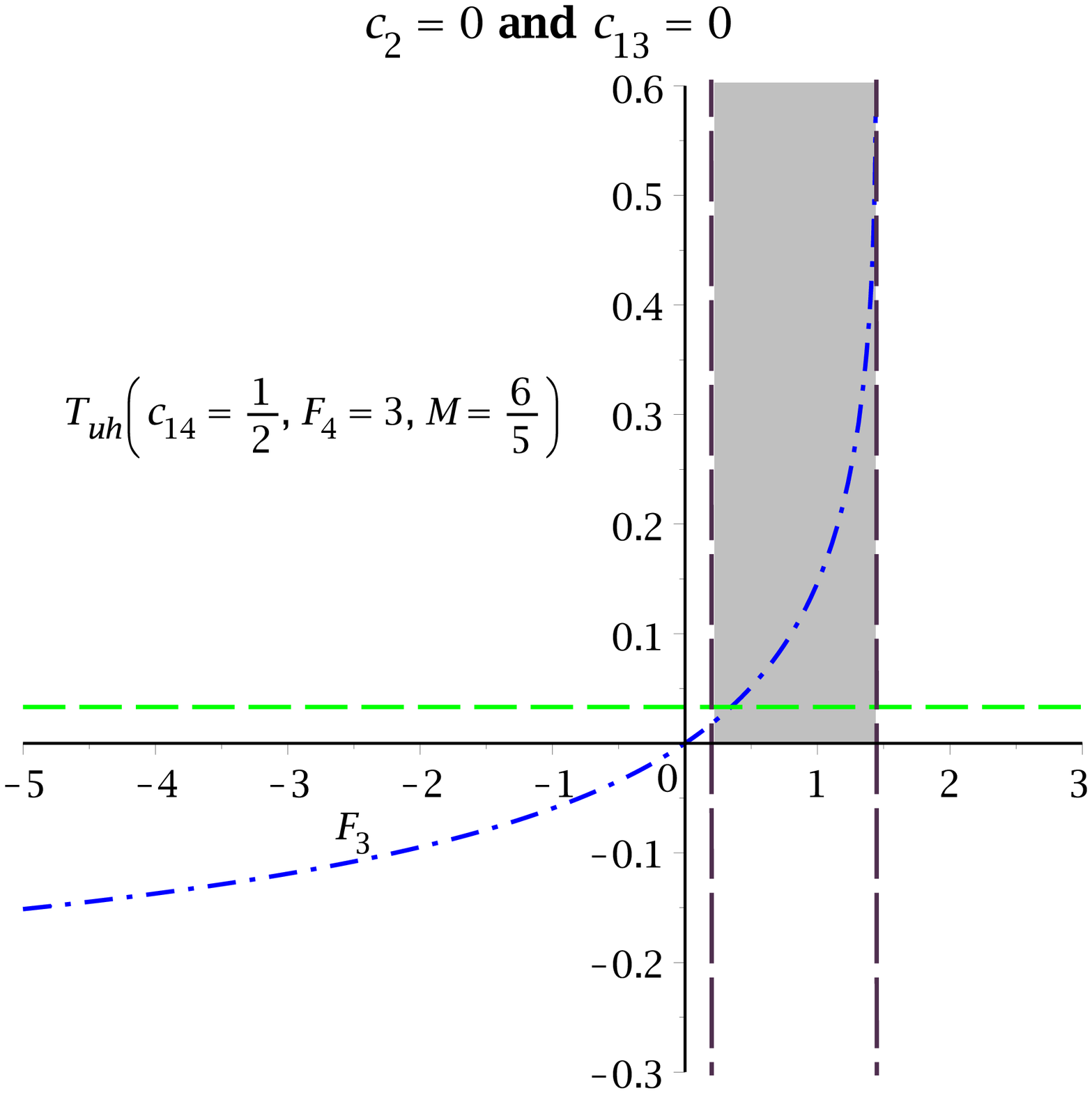}
	\includegraphics[width=6.5cm]{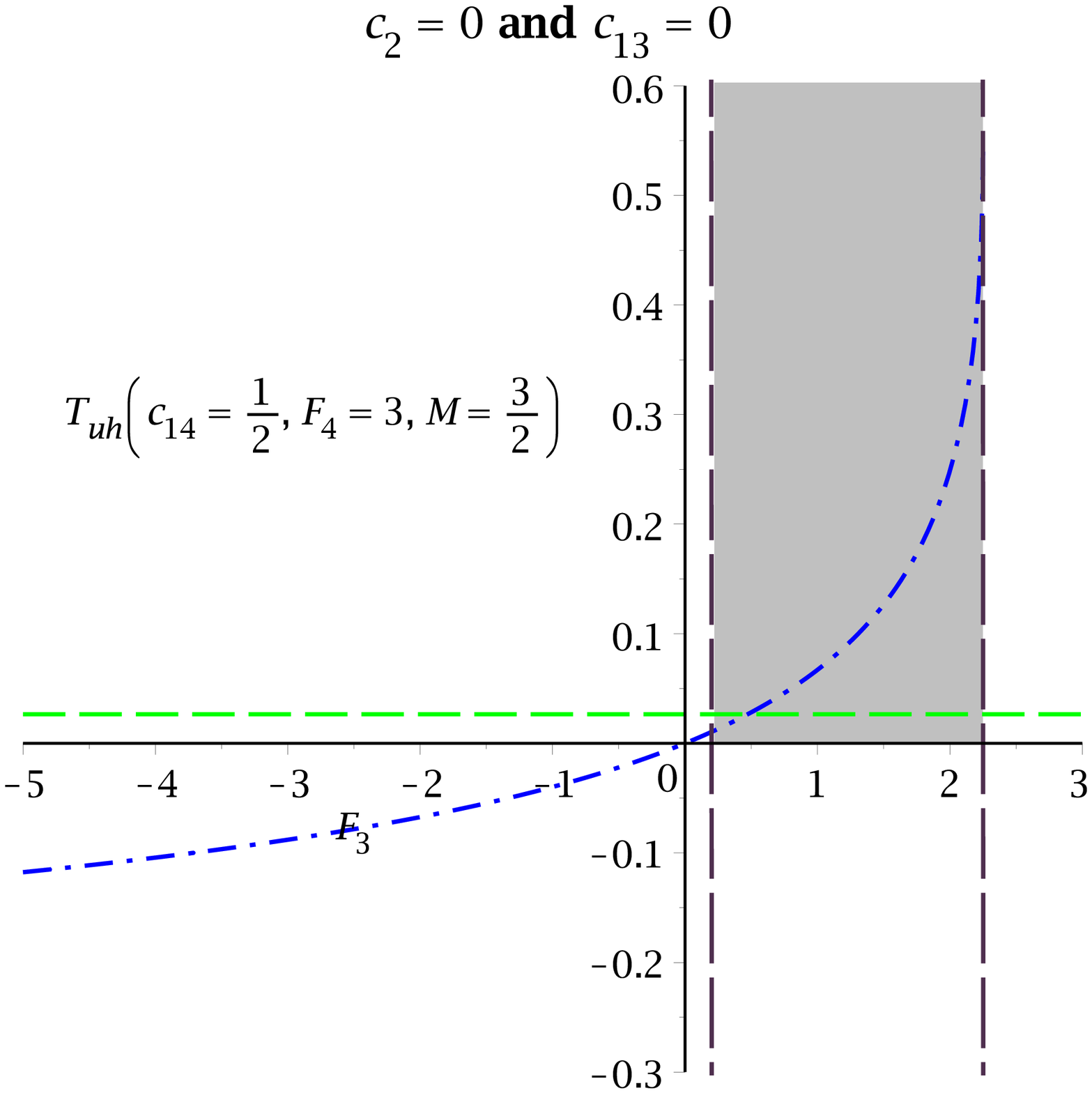}
	\includegraphics[width=6.5cm]{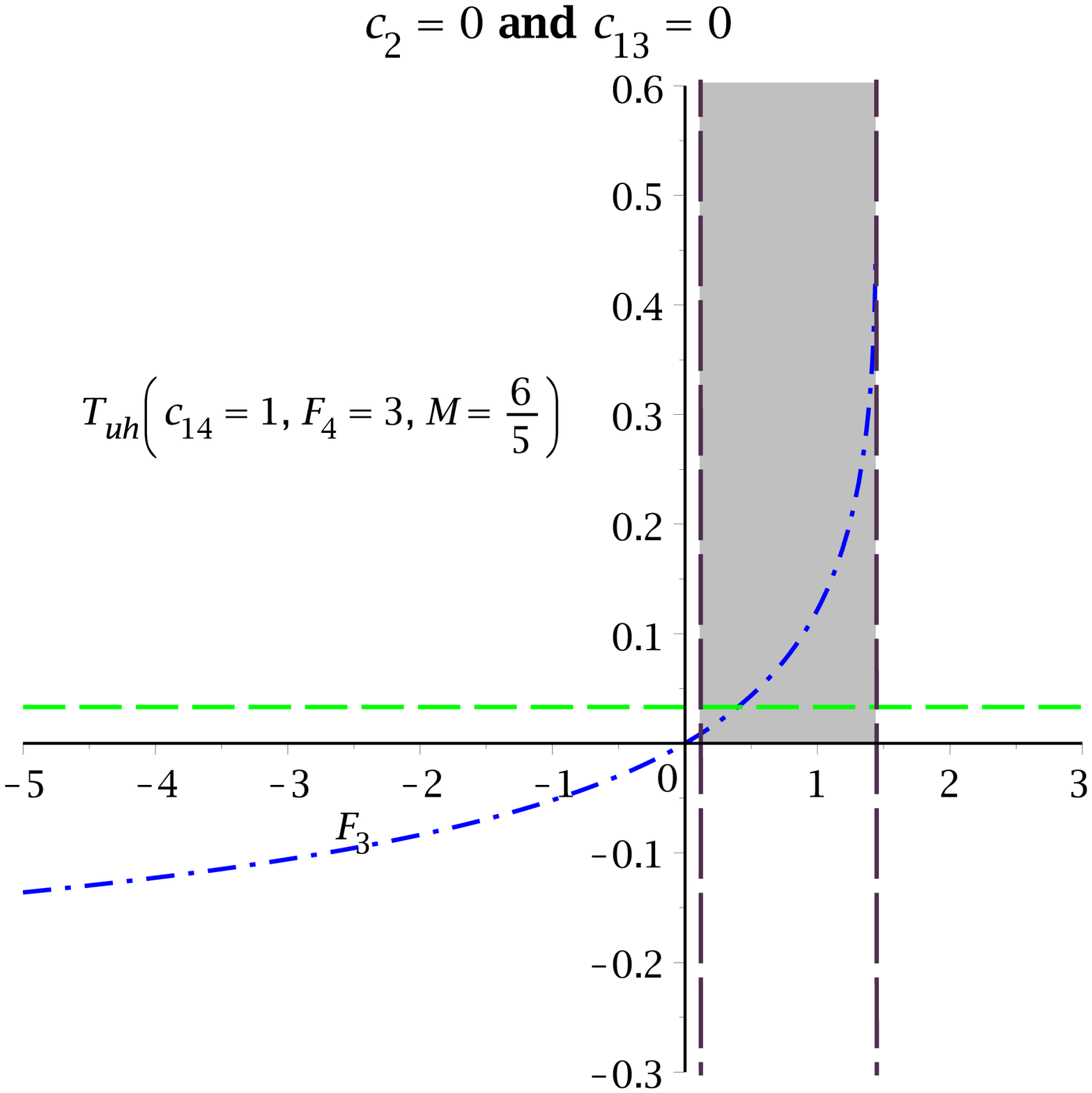}
	\includegraphics[width=6.5cm]{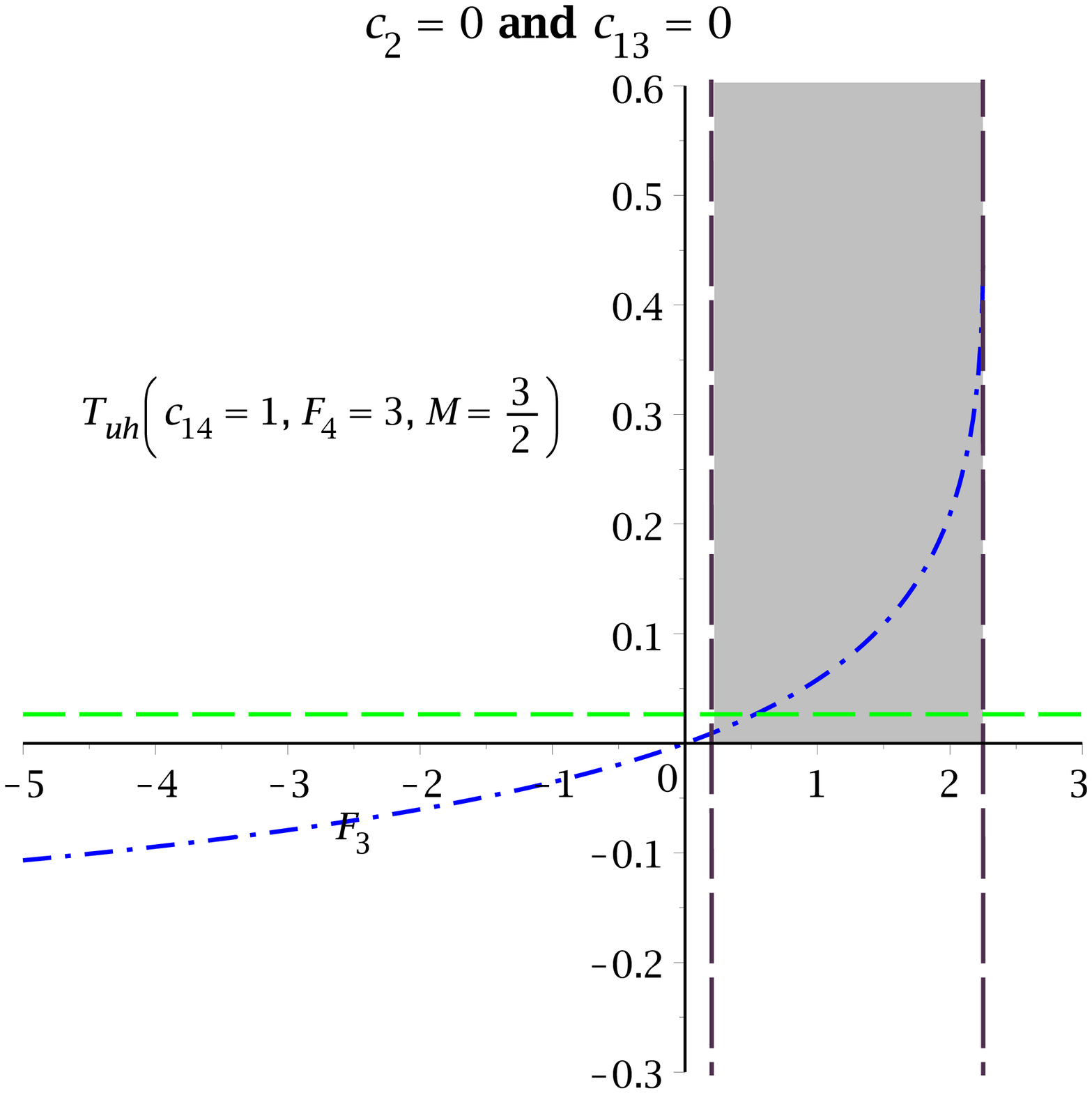}
	\includegraphics[width=6.5cm]{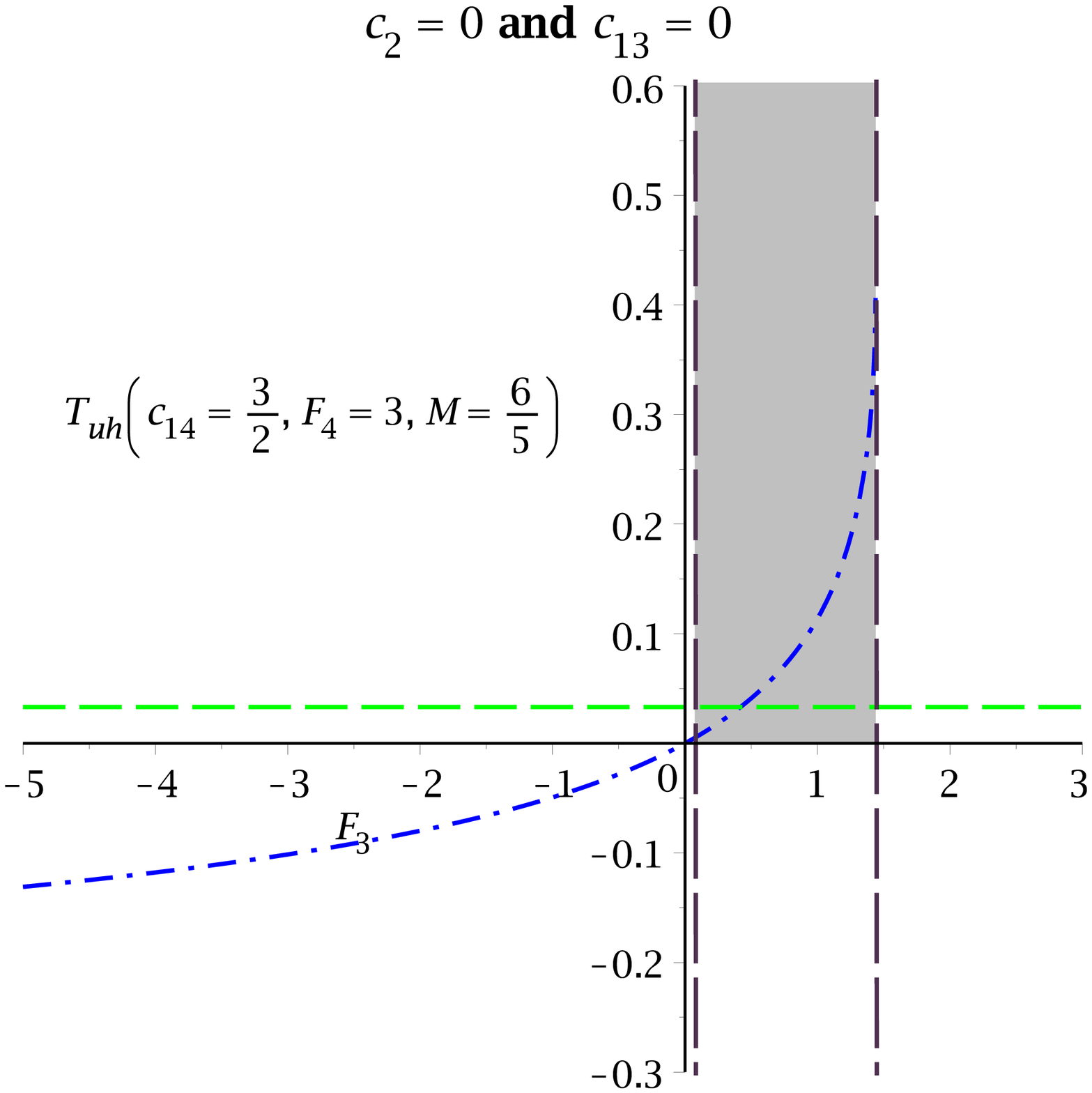}
	\includegraphics[width=6.5cm]{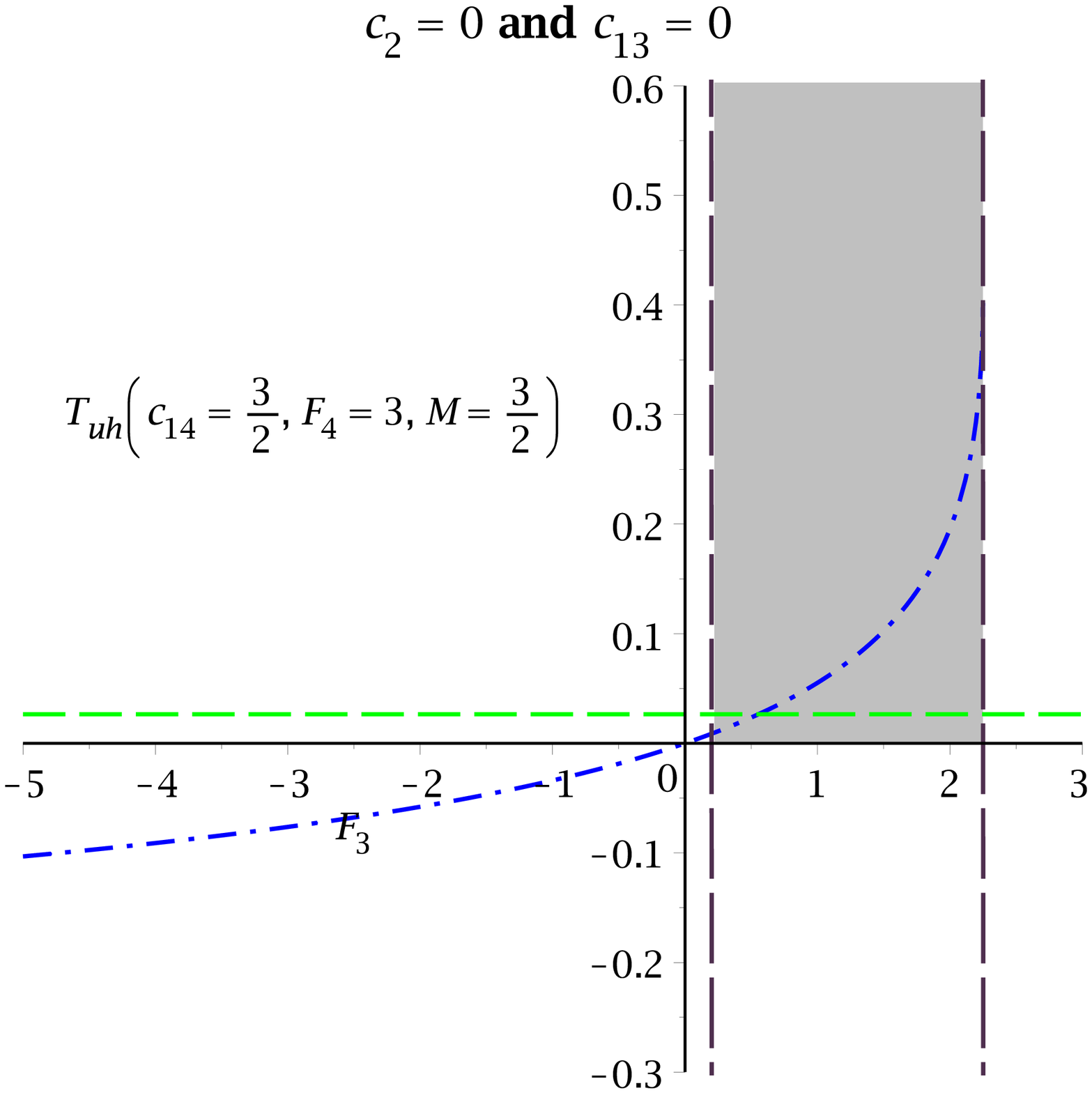}
\caption{These figures show the universal horizon temperatures, 
for the Case \textcolor{black}{D}, where we have: $T_{uh1}$ (blue dot-dashed line) and
$T_{GR}$ (green dashed line). 
	The violet long-dashed straight lines
	represent the inferior and superior limits of $F_3$. 
	The gray areas are the regions where the condition (\ref{cond1}) is valid.}
	\label{figT12}	
\end{figure}

As we saw earlier, $F_4$ must be positive leading to $r_{uh1}<0$ and therefore this does not correspond to a real horizon. Thus, we have that $r_{uh2,3}$ are the outermost universal horizon, 
if $F_3 \le M^2$, and
we have for the surface gravity, temperature, entropy and the first law, 
using equations (\ref{termu})
\textcolor{black}{
\bqn
\kappa_{uh2} &=& {{\frac{{F_3} \left[ 2 + \sqrt {2}{c_{14}} {F_4} \left( M + \sqrt{{M}^{2}-{F_3}} \right) \right] }{ 2 {c_{14}} \left( M+\sqrt {{M}^{2}-{F_3}} \right)^{3}}}} \\
T_{uh2} &=& {{\frac{{F_3} \left[ 2 + \sqrt {2}{c_{14}} {F_4} \left( M + \sqrt{{M}^{2}-{F_3}} \right) \right] }{ 4 \pi {c_{14}} \left( M+\sqrt {{M}^{2}-{F_3}} \right)^{3}}}} \\
S_{uh2} &=& {\frac {\pi   \left( M+\sqrt {{M}^{2}-{F_3}} \right) ^{2}}{G}},\\
\delta S_{uh2} &=&  \frac{ 4 \pi {c_{14}} \left( M+\sqrt {{M}^{2}-{F_3}} \right)^{3} \delta M}{{F_3} \left[ 2 + \sqrt {2}{c_{14}} {F_4} \left( M + \sqrt{{M}^{2}-{F_3}} \right) \right] }.
\eqn
}
Then, when we establish the condition $F_3 < M^2$, this is in agreement with the solution of the Reissner-Nordstr\"{o}m metric.
See Figures \ref{figC1}, \ref{figC2} and \ref{figT12}.
From the figures \ref{figC1}, \ref{figC2} and \ref{figT12}, 
as already pointed out, we can see that for values of $F_3>M^2$ we do
not have any horizon, thus, we have naked singularities.

\clearpage

\textcolor{black}{
\section{\bf Solutions for case (E): \textcolor{black}{$\bf c_{2} =-c_{13} \neq 0$ and $\bf c_{14} \neq 0$}}
%
The solution of the field equations (\ref{Gtt})-(\ref{Gphiphi}) for 
$c_{2}=-c_{13}$ we get, 
\bqn
&&A=H_2+\frac{H_4}{r}+\frac{H_3}{r^2},\nb\\
&&B=\frac{H_1}{H_2+\frac{H_4}{r}+\frac{H_3}{r^2}},\nb\\
\eqn
where $H_1$, $H_2$, $H_3$ and $H_4$ are arbitrary integration constants
we have chosen $H_1=H_2=1$ in order to have
a flat spacetime at infinity and $H_4=-2M$ in order to have a resemblance
with the Schwarzschild solution, where $M$ is the 
Schwarzschild mass. 
Thus, $A(r)$ and $B(r)$ can be rewritten as
\bqn
&&A=1-\frac{2M}{r}+\frac{H_3}{r^2},\nb\\
&&B=\frac{1}{1-\frac{2M}{r}+\frac{H_3}{r^2}}.
\eqn
The solutions for $a(r)$ and $b(r)$ are
\bqn
&&a =  \frac{\epsilon}{r} {\sqrt {{\frac {2\sqrt{2} \zeta {r} \Delta + 2Mr \left( 2 {c_{13}} - {c_{14}} \right) + {H_3}\left( {c_{14}} - 2 \right) }{\left( 2 {c_{13}} - {c_{14}} \right)}}}}
\nb\\
&&b =\frac{\epsilon r \left( \sqrt{2 H_3 (- 1 - {c_{13}})} + \zeta r \sqrt{(2{c_{13}} - {c_{14}})} \right) }{\left( 2 Mr -{r}^{2}-{H_3} \right) \sqrt{(2{c_{13}} - {c_{14}})}},
\eqn
where ${\bf \Delta}=\sqrt {{H_3}  \left( -1+{c_{13}} \right)  \left( 2 {c_{13}}-{c_{14}}  \right) }$ and ${H_3}  \left( -1+{c_{13}} \right)  \left( 2 {c_{13}}-{c_{14}}
 \right)  > 0$, in order to ensure that the components of the aether vector are real. 
Note that the solutions presented in this section depend explicitly on the parameters 
$c_{13}$ and $c_{14}$.
The Kretschmann scalar is given by
\bq
K = {\frac {8(7 {{H_3}}^{2}-12 {H_3} Mr+6 {M}^{2}{r}^{2})}{{
r}^{8}}}.
\eq
Note that $r=0$ is the singularity of the spacetime.
The Killing horizon equation is given by
\bqn
&&{\chi}^{\alpha} {\chi}_{\alpha} = -1+\frac{2M}{r}-\frac{H_3}{r^2}=0,\nb\\
\eqn
whose roots are
\bqn
&&r_{kh1,2} = M+ \eta_1 \sqrt{M^2- H_3},
\eqn
where $H_3 \le M^2$.
The universal horizon equation is
\bqn
&&{\chi}^{\alpha} {u}_{\alpha} = \sqrt {-{\frac {2 \sqrt {2}{\bf \Delta}+
2 {c_{13}} {H_3}-2 {r}^{2}{{c_{13}}}^{2}-2 {{c_{13}}}^{2}{
H_3}+{c_{13}} {c_{14}} {r}^{2}}{{r}^{2} \left( 2 {c_{13}}-{
c_{14}} \right) {c_{13}}}}}, 
\lb{uhE}
\eqn
whose four solutions are
\bqn
&&r_{uh1,2} = \eta_1 {{\sqrt{ \frac {{2}{H_3}  \left( -1+{c_{13}} \right) }{2 {c_{13}}-{c_{14}}}}}},
\eqn
%
Since the outermost universal horizon is $r_{uh1}$, the surface gravity, temperature, entropy and the first law and
using equations (\ref{termu}) we have
\bqn
\kappa_{uh1} &=& \frac{\sqrt {2}}{8} \times \nb\\
&&\sqrt {{\frac { \left( -2 {c_{13}} {H_3}+2 {
c_{13}} M\sqrt {2}{\bf \Delta}_1+{c_{14}} {H_3} {c_{13}}+4  \left| {c_{13}}  \left( -1+{c_{13}} \right) {H_3}
 \right|  \right) {c_{13}}}{ \left( -1+{c_{13}} \right) {H_3}}}
}\times \nb\\
&&\sqrt {{{c_{13}}}^{2}{H_3}-{c_{13}} {H_3}+ \left| {c_{13}}  \left( -1+{c_{13}} \right) {H_3} \right| } \left| {\frac 
{2 {c_{13}}-{c_{14}}}{{c_{13}}  \left( -1+{c_{13}} \right) {H_3}}} \right| \times\nb\\
&&{\frac {1}{\sqrt {{c_{13}}  \left( 2 {
c_{13}}-{c_{14}} \right) }}}
, \\
T_{uh1} &=& \frac{\sqrt {2}}{16{\pi } }\times \nb\\
&&\sqrt {{\frac { \left( -2 {c_{13}} {H_3}+2 {
c_{13}} M\sqrt {2}{\bf \Delta}_1+{c_{14}} {H_3} {c_{13}}+4  \left| {c_{13}}  \left( -1+{c_{13}} \right) {H_3}
 \right|  \right) {c_{13}}}{ \left( -1+{c_{13}} \right) {H_3}}}
}\times\nb\\
&&\sqrt {{{c_{13}}}^{2}{H_3}-{c_{13}} {H_3}+ \left| {c_{13}}  \left( -1+{c_{13}} \right) {H_3} \right| } \left| {\frac 
{2 {c_{13}}-{c_{14}}}{{c_{13}}  \left( -1+{c_{13}} \right) {H_3}}} \right| \times\nb\\
&&{\frac {1}{\sqrt {{c_{13}}  \left( 2 {
c_{13}}-{c_{14}} \right) }}}
,\\
S_{uh1} &=& {\frac {2 \pi {H_3}  \left( -1+{c_{13}} \right) }{ G \left( 2 {
c_{13}}-{c_{14}} \right) }}
,\\
\delta S_{uh1} &=& 8\pi\, \delta M\, \sqrt {2} \left| {\frac {{c_{13}}  \left( -1+{c_{13}}
 \right) {H_3}}{2 {c_{13}}-{c_{14}}}} \right| \sqrt {{
c_{13}}  \left( 2 {c_{13}}-{c_{14}} \right) }\times\nb\\
&&{\frac {1}{\sqrt {{
\frac { \left( -2 {c_{13}} {H_3}+2 {c_{13}} M\sqrt {2}
{\bf \Delta}_1+{c_{14}} {H_3} {c_{13}}+4  \left| {c_{13}}  \left( -1+{c_{13}} \right) {H_3} \right|  \right) {c_{13}}}{ \left( -1+{c_{13}} \right) {H_3}}}}}}\times
\nb\\
&&{\frac {1}{\sqrt {{{
c_{13}}}^{2}{H_3}-{c_{13}} {H_3}+ \left| {c_{13}} 
 \left( -1+{c_{13}} \right) {H_3} \right| }}},
\eqn
where $\epsilon=\zeta=1$.
%
\section{{\bf Solutions for case (F):} $\bf c_{2} = 0$  and $\bf c_{13} \neq 0$ and $\bf c_{14} \neq 0$}
%
The solution of the field equations (\ref{Gtt})-(\ref{Gphiphi}) for $c_{2}=0$ is given by $a(r)=0$, i.e., the radial component of the aether vector is null,
thus, we have a static aether. Therefore, we have not considered this case.
%
\section{{\bf Solutions for case (G):} $\bf c_{2} \neq 0$  and $\bf c_{13} = 0$ and $\bf c_{14} \neq 0$}
%
The solution of the field equations (\ref{Gtt})-(\ref{Gphiphi}) for 
$c_{13}=0$ it is also given by $a(r)=0$, i.e., the radial component of the aether vector is null,
thus, again we have a static aether. Therefore, we have not considered this case.
%
\section{{\bf Solutions for case (H):} $\bf c_{2} = -c_{13} \neq 0$ and $\bf c_{14} = 0$}
%
The solution of the field equations (\ref{Gtt})-(\ref{Gphiphi}) for 
Thus, $A(r)$ and $B(r)$ can be rewritten as
\bqn
&&A={\frac {{J_2}}{r}},\nb\\
&&B={J_1}\,r.
\eqn
The solutions for $a(r)$ and $b(r)$ are
\bqn
&&a ={\frac {\sqrt {{c_{13}}\,r \left( r+{J_3}\,{c_{13}} \right) }}{
{c_{13}}\,r}},\nb\\
&&b =\sqrt {{\frac {r\left( {J_1\,}r+{J_1}\,{J_3\,
}{c_{13}}+{c_{13}} \right) }{{J_2\,}{c_{13}}}}},
\eqn
where $J_1$, $J_2$ and $J_3$ are arbitrary integration constants.
This solution does not have a flat spacetime at infinity.
Therefore, we have also not considered this case.
}

\section{Conclusions}

In the present work, we analyze several spherically symmetric exterior vacuum solutions allowed by the Einstein-Aether (EA) theory with a non-static aether. \textcolor{black}{We show that there are five classes of solutions corresponding to different values of a combination of the free parameters, $c_{2}$, $c_{13}=c_1+c_3$ and $c_{14}=c_1+c_4$, which are: (A) $c_{14}=0$ (B) $c_{13}=0$ and $c_{14}=0$, (C) $c_2=0$ and $c_{14}=0$, (D) $c_2=0$ and $c_{13}=0$, and (E) $c_2=- c_{13}\neq 0$. We present explicit analytical solutions for these five cases. The cases where only $c_2=0$ or only $c_{13}=0$ are not analytic solutions. All these cases present singularities at $r=0$ and are asymptotically flat spacetimes,  and posses both Killing, and universal horizons.} We call attention to the fact that in all the cases presented here, we
have several solutions of the aether vector field for the same spacetime. This means that
the geometry of the spacetime, defined by the metric, is not sensitive to different aether fields of the same spacetime.
Also, it should be noted that all the solutions presented in this paper depend explicitly on some of the aether parameters 
except the solution \textcolor{black}{(B)}.

\textcolor{black}{
We have shown that the universal horizons are always situated inside than the Killing horizons. We have also computed the surface gravity, the temperature, the entropy, and the first law of thermodynamics for the outermost universal horizons. The temperature of a black hole in EA theory can be higher or lower than in GR. In Case (B), the temperature is always lower than GR. However, in the Cases \textcolor{black}{(A), (C), (D) and (E)} depend on the values of $c_{13}$, $c_{14}$ or $c_2$. Besides, the Case (D) also depends on two arbitrary constants ($F_3$ and $F_4$) [see Table \ref{table4}].
We also notice that the temperature tends to $+\infty$ when $c_{13} \rightarrow 1$ in Cases
(C) and (E), and when $F_3 \rightarrow M^2$ in Case (D) and also when $c_{13} \rightarrow -3c_2$ in the Case (E). See Figure \ref{figT12} for the details.} 

\textcolor{black}{
As with temperature, the entropy of the EA black hole can also be higher or lower than in GR. In Cases (A), (B), and (C) the entropy is lower than the GR. However, in Case (D)} this quantity depends on the values of $c_{14}$ and of the arbitrary constant $F_3$, with, $c_{14}$ causing entropy to increase and $F_3$ causing it to decrease, in comparison with the GR  (see Table \ref{table4}).
The lowest value of the entropy is for the values $c_{14}=0$ and $F_3=M^2$, giving $S=\pi M^2$. 
We can also notice that the entropy tends to $+\infty$ when $c_{14} \rightarrow 2$ in this case.

\begin{table*}
\centering
\begin{minipage}{175 mm}
\caption{Summary of the Solutions}
\label{table3}
\begin{tabular}{@{}|c|l|l|c|c|}

\hline
Case & Metric Functions & Aether Components & Horizons & Obs \\

\hline
\multirow{3}{*}{$\textcolor{black}{(A)}$} & $A=1-\frac{2M}{r}+\frac{G_4}{r^4}$ & 
$a=\frac{-3\sqrt{3} \zeta {M}^{2}}{4 r^2 \sqrt {1 -{c_{13}}}}$  &  & (1) \\
 & \textcolor{black}{$B=\left({1-\frac{2M}{r}+\frac{G_4}{r^4}}\right)^{-1}$} &  & $r_{uh}=\frac{3M}{2}$ & \\
 &  $G_4= \frac{27}{16} \left( \frac{c_{13}}{c_{13}-1} \right) M^4$ & 
 $b= \frac{ 4\epsilon \left( {c_{13}}-1 \right) \left( -2r+3M \right) {r}^{2} \sqrt{ 3{M}^{2}+4Mr+4{r}^{2}}} {\left( {c_{13}}-1 \right) 16{r}^{3} \left(  r - 2m \right)+27{c_{13}}{M}^{4}}$ & & \\
 
\hline
\multirow{2}{*}{$\textcolor{black}{(B)}$} & $A=1-\frac{2M}{r}$  & 
$a={\frac {3 \sqrt {3}{M}^{2}}{4{r}^{2}}}$ 	 & \textcolor{black}{$r_{uh}=\frac{3M}{2}$} & (2) \\
 & $B=\left( 1-\frac{2M}{r} \right)^{-1}$  & 
 $b={\frac{ \epsilon \left( -2r+3M \right) \sqrt{3{M}^{2}+4Mr+4{r}^{2}} }{4{r} \left( -r+2M \right)}}$ &  & \\
		
\hline
\multirow{3}{*}{$\textcolor{black}{(C)}$} & $A=1+\frac{E_2}{r^4}-\frac{2M}{r}$ & 
$a=\frac{3\sqrt{3} \zeta {M}^{2}}{4 r^2 \sqrt {1 -{c_{13}}}}$ &  &  \\
 & $B=\left({1+\frac{E_2}{r^4}-\frac{2M}{r}}\right)^{-1}$ & 
  &  $r_{uh}=\frac{3M}{2}$   &  \\
 & $E_2 = \frac{27}{16} \left( \frac{c_{13}}{c_{13}-1} \right) M^4$ & 
 $b= \frac{ 4\epsilon \left( {c_{13}}-1 \right) \left( -2r+3M \right) {r}^{2} \sqrt{ 3{M}^{2}+4Mr+4{r}^{2}}} {\left( {c_{13}}-1 \right) 16{r}^{3} \left(  r - 2m \right)+27{c_{13}}{M}^{4}}$ & & \\	

\hline
\multirow{3}{*}{$\textcolor{black}{(D)}$} & $A=1-\frac{2M}{r}+\frac{F_3}{r^2}$ & \textcolor{black}{$a=\frac{\zeta}{r} \sqrt{ \frac{F_3(c_{14} F_4 r + \sqrt{2})^2-c_{14}(r^2-2M r+F_3)}{c_{14}}}$} & $r_{uh}=M+$ & \\
 & \textcolor{black}{$B=\left({1-\frac{2M}{r}+\frac{F_3}{r^2}}\right)^{-1}$} & $b=\frac{\epsilon\; r (c_{14} F_4 r + \sqrt{2}) }{(r^2-2M r+F_3)} \sqrt{\frac{F_3}{c_{14}}}$ & $\sqrt{ M^2- F_3}$ & 
 \\
 & $\frac{1}{{F_4}^2 c_{14}} < F_3 \le M^2 $ & $F_4>0$ &  & \\	
\cline{2-5}
\multirow{3}{*}{} & \textcolor{black}{$A=1-\frac{2M}{r}+\frac{M^2}{r^2}$} & \textcolor{black}{$a=\frac{\zeta}{r} \sqrt{ \frac{M^2(c_{14} F_4 r + \sqrt{2})^2-c_{14}(r^2-2M r+M^2)}{c_{14}}}$} & \textcolor{black}{$r_{uh}=M$} & \\
 & \textcolor{black}{$B=\left({1-\frac{2M}{r}+\frac{M^2}{r^2}}\right)^{-1}$} & \textcolor{black}{$b=\frac{\epsilon\; r (c_{14} F_4 r + \sqrt{2}) }{(r^2-2M r+M^2)} \sqrt{\frac{M^2}{c_{14}}}$} &  & \\
 & \textcolor{black}{$F_3=M^2$} &  &  & \\	
 
\hline
\multirow{3}{*}{\textcolor{black}{$(E)$}} & \textcolor{black}{$A=1-\frac{2M}{r}+\frac{H_3}{r^2}$} & \textcolor{black}{$a= \frac{\epsilon}{r} {\sqrt {{\frac {2\sqrt{2} \zeta {r} \Delta + 2Mr \left( 2 {c_{13}} - {c_{14}} \right) + {H_3}\left( {c_{14}} - 2 \right) }{\left( 2 {c_{13}} - {c_{14}} \right)}}}}$} & \textcolor{black}{$r_{uh}=$} & \\
 & \textcolor{black}{$B=\left({1-\frac{2M}{r}+\frac{H_3}{r^2}}\right)^{-1}$} &  &  \textcolor{black}{$\sqrt {\frac{ {2}{H_3} ( -1+{c_{13}} ) }{( 2{c_{13}}-{c_{14}})}}$} & \\
 &  \textcolor{black}{${
H_3}  \left( -1+{c_{13}} \right) \times$} & \textcolor{black}{$b =\frac{\epsilon r \left( \sqrt{2 H_3 (- 1 - {c_{13}})} + \zeta r \sqrt{(2{c_{13}} - {c_{14}})} \right) }{\left( 2 Mr -{r}^{2}-{H_3} \right) \sqrt{(2{c_{13}} - {c_{14}})}}$} &  &\\
 & \textcolor{black}{$\left( 2 {c_{13}}-{c_{14}}\right)  > 0$} &  & & \\
 & \textcolor{black}{$\&$ $H_3\leq M^2$} & \textcolor{black}{${\bf \Delta}=\sqrt {{
H_3}  \left( -1+{c_{13}} \right)  \left( 2 {c_{13}}-{c_{14}}
 \right) }$} &  & \\	
\cline{2-5}
\multirow{3}{*}  & \textcolor{black}{$A=1-\frac{2M}{r}+\frac{M^2}{r^2}$} & \textcolor{black}{$a= \frac{\epsilon }{r} \sqrt{\frac{(\sqrt{2M^2}-r\sqrt{c_{14}})^2-c_{14}(r^2-2Mr+M^2)}{c_{14}}}$} & \textcolor{black}{$r_{uh}=M\sqrt{\frac{2}{c_{14}}}$} & \\
 & \textcolor{black}{$B=\left({1-\frac{2M}{r}+\frac{M^2}{r^2}}\right)^{-1}$} & \textcolor{black}{$b =\frac{\epsilon r }{2Mr-r^2-M^2} \left( \sqrt{\frac{2M^2}{c_{14}}} +r \right)$} & & \\
 &  \textcolor{black}{$c_{13}=0$ $\&$ $H_3= M^2$} & & & \\





\hline
\end{tabular}
		
\medskip
\textcolor{black}{
Notes: (1) Notice that $G_4 \equiv E_2$ of the Case (C).
(2) $c_2 \ne 0$, yet black hole thermodynamics is not exactly the same as in GR, where $r_{uh} = 2M$.}
\end{minipage}

\end{table*}

\clearpage 

\begin{table*}
\centering
\begin{minipage}{160 mm}
\caption{Summary of the Temperature and Entropy}
\label{table4}
\begin{tabular}{@{}|c|l|c|c|}

\hline
Case & Temperature & Entropy & Notes\\

\hline
\multirow{1}{*}{$\textcolor{black}{(A)}$} & $\frac{\sqrt {6}}{18\pi M \sqrt {1-{c_{13}}} } $  & $\frac{9\pi M^2}{4}$ & \\
 & & & \\
 
\hline
\multirow{1}{*}{$\textcolor{black}{(B)}$} & \textcolor{black}{${\frac {\sqrt {6}}{18M\pi }}$} & \textcolor{black}{${\frac {9\pi  {M}^{2}}{4}}$} & \\
 & & & \\

\hline
\multirow{1}{*}{$\textcolor{black}{(C)}$} & \textcolor{black}{${\frac {\sqrt {6}}{18\pi M\sqrt {1-{c_{13}}} }}$} & \textcolor{black}{${\frac {9\pi  {M}^{2}}{4}}$} &   \\
 & & & \\
 
\hline
\multirow{1}{*}{$\textcolor{black}{(D)}$} 
& ${{{{\frac{{F_3} \left[ 2 + \sqrt {2}{c_{14}} {F_4} \left( M + \sqrt{{M}^{2}-{F_3}} \right) \right] }{ 4 \pi {c_{14}} \left( M+\sqrt {{M}^{2}-{F_3}} \right)^{3}}}}}}$
 & $\frac {2\pi \left( M+\sqrt {{M}^{2}-{F_3}} \right) ^{2}}{2-c_{14}}$ & (1) \\
 &   & &  \\
\cline{2-4} 
\multirow{1}{*}  & \textcolor{black}{${\frac {2+\sqrt {2}{F_4} {c_{14}}M}{4 \pi c_{14} M}}$} & \textcolor{black}{$\frac {2\pi M^{2}}{2-c_{14}}$} & \textcolor{black}{(2)} \\
 & & & \\

\hline
\multirow{1}{*}{$\textcolor{black}{(E)}$} 
& \textcolor{black}{$\frac{1}{8{\pi } }  \left| {\frac{\sqrt {2 {c_{13}}-{c_{14}}}}{ \left( -1+{c_{13}} \right) {H_3}}} \right| \times$} 
& \textcolor{black}{${\frac {2 \pi {H_3} \left( -1+{c_{13}} \right) }{ G \left( 2 {c_{13}}-{c_{14}} \right) }}$} 
& \textcolor{black}{(3)} 
\\ 
& \textcolor{black}{$\sqrt{{ 2 \sqrt {2}  M {\bf \Delta} + {H_3} \left( {c_{14}}  -2 \right) +4 {H_3} \left( -1+{c_{13}} \right) }}$} 
& 
&  
\\
\cline{2-4} 
\multirow{1}{*}  & \textcolor{black}{${\frac{\sqrt{{c_{14}} \left( 6 - {c_{14}} - 2 \sqrt{ 2{c_{14}}} \right)}}{8 \pi}}$} & \textcolor{black}{$\frac {2\pi M^{2}}{c_{14}G}$} & \textcolor{black}{(4)} \\
 & & & \\


\hline
\end{tabular}
		
\medskip
Notes:  (1) 
See also the Figure \ref{figT12}. Notice that $G$ for the Cases (A), (B) and (C) is
equal to $G_N=1$, since $c_{14}=0$.
\textcolor{black}{(2) Solution (D) assuming $F_3=M^2$. 
(3) ${\bf \Delta}=\sqrt { \left( 2 {c_{13}}-{c_{14}} \right) \left( -1+{c_{13}} \right) {H_3}}$ (4) assuming $H_3=M^2$ and $c_{13}=0$}
\end{minipage}

\end{table*}


\textcolor{black}{
Finally, we want compare our results in Schwarzschild coordinates with those  of the references \cite{Ding:2015kba} \cite{Zhang:2020too} \cite{Berglund2012}
\cite{Bhattacharyya2014} in Eddington-Finkelstein coordinates, and show that our results are new. In order to compare, we first transform their results in Eddington-Finkelstein coordinates into Schwarzschild coordinates.
From the Eddington-Finkelstein metric, we have
\bq
ds^2= -e(r) dv^2+2f(r) dr dv +r^2 d\theta^2 +r^2 \sin^2 \theta d\phi^2,
\eq
with the aether vector given by
\bq
u^a= \left[ \alpha_1(r), \beta_1(r), 0, 0 \right],
\eq
when normalized we get
\bq
u^a= \left[ \alpha_1(r), \frac{e(r)\alpha_1(r)^2-1}{2f(r)\alpha_1(r)}, 0, 0 \right].
\eq
We can make a coordinate transformation ($dv=dt+dr/e(r)$ and assuming the same radial coordinate) 
in order to transform them into Schwarzschild 
coordinates (see more details in \cite{Zhu2019}), thus we get
\bq
ds^2= -e(r) dt^2+e(r)^{-1} dr^2 +r^2 d\theta^2 +r^2 \sin^2 \theta d\phi^2,
\eq
with the aether vector given by
\bq
u^a= \left[ \alpha_1(r) -\frac{\beta_1(r)}{e(r)}, \beta_1(r), 0, 0 \right],
\eq
when normalized we obtain
\bqn
u^a= \left[ \frac{\alpha_1(r) e(r)-\beta_1(r)}{e(r) \sqrt{\alpha_1(r) [\alpha_1(r) e(r)-2 \beta_1(r)]}}, \frac{\beta_1(r)}{\sqrt{\alpha_1(r) [\alpha_1(r) e(r)-2 \beta_1(r)]}}, 0, 0 \right],\nb\\
\eqn
and the timelike Killing vector is also given by equation (\ref{chia}).
In these previous papers, they have presented only two analytical solutions for $c_{14}=0$ 
(Case I) and $c_{123}=0$ (Case II). Using these coordinate transformations we can
get their results in our coordinates. See Table \ref{table6} for the details.
Comparing the Tables \ref{table3} and \ref{table4} with the correspondent cases in Table \ref{table6}, we observe 
clearly from the metric functions, the aether vectors
 and the universal horizons that they are different from each other. However, the universal horizons 
 of the Cases (A), (B) and (C) coincide with the Case I.
%
We notice
that the thermodynamical quantities such as temperature and entropy, of the Case I coincide with
ours ones of the Case A. In the rest of the cases the thermodynamical quantities are 
different to each other.
Thus, we conclude that our results are completely different from
the previously published papers (albeit in a different coordinate system), except our Case (C) coincides with Case I . 
The reason that the universal horizons  and their thermodynamical properties are different is because the surface gravity depends explicitly on the aether vector. 
%
%
%
\begin{table*}
\centering
\begin{minipage}{160 mm}
\caption{\textcolor{black}{Summary of the results of Eddington-Finkelstein coordinates 
transformed into  Schwarzschild coordinates \cite{Ding:2015kba}\cite{Zhang:2020too}\cite{Berglund2012}\cite{Bhattacharyya2014}}}
\label{table6}
\begin{tabular}{@{}|l|l|l|}
\hline
Cases & I ($c_{14}=0$) & II ($c_{123}=0$) \\
\hline
\multirow{2}{*}{Metric Functions} & 
$A=1-\frac{2M}{r}-\frac{E_2}{r^4}$  & 
$A=1-\frac{2M}{r}+\frac{r_u(2M+r_u)}{r^2}$ \\
                 & 
$B=\frac{1}{A}$ & 
$B=\frac{1}{A}$  \\
\hline
\multirow{2}{*}{Aether Components} & 
$a = -\frac{\sqrt{E_2}}{r^2\sqrt{c_{13}}}$  & 
$a = -\frac{M+r_u}{r}$   \\
 & 
$b = \frac{\sqrt{r^4[c_{13}( r^4-2 M r^3)+E_2(1-c_{13})]}}{\sqrt{c_{13}}(r^4-2 r^3 M-E_2)}$ &
$b = \frac{(r-M) r}{r^2+2 M r-2 r_u M-r_u^2}$ \\
\hline
\multirow{1}{*}{Universal Horizon} & 
$r_{uh}=\frac{3M}{2}$ &
$r_{uh}=M$ \\
\hline
\multirow{1}{*}{Temperature} & 
$T = {\frac {\sqrt {6}}{18M\pi  \sqrt {1-{c_{13}}}}}$ & 
$T = {\frac {\sqrt {2}\sqrt {2-{c_{14}}}}{8M\pi }}$ \\
\hline
\multirow{1}{*}{Entropy} & 
$S = {\frac {9\pi  {M}^{2}}{4}}$ &
$S = \frac{2\pi M^2}{2-c_{14}}$ \\
\hline
\end{tabular}
\medskip
\medskip
\\
\textcolor{black}{
Note that, using the notation of the references \cite{Berglund2012}
\cite{Bhattacharyya2014}, we have the following relations: $e(r) \equiv A$;\,
$B(r) \equiv e(r)^{-1}$;\, $r_0 \equiv 2M$;\, 
$ r_{\ae}^4=\frac{27}{16} \left( \frac{c_{13}}{1-c_{13}} \right) M^4 \equiv \frac{E_2}{c_{13}}$;\,
$r_u =  M \sqrt{1-\frac{c_{14}}{2}}$, that help to make the comparison. In reference
\cite{Ding:2015kba}, the authors assume null charge. The universal horizons are reobtained from the
the transformed metrics and aether vectors.}
\end{minipage}
\end{table*}
}
%

\section {Acknowledgments}
We would like to thank Dr. Anzhong Wang for valuable suggestions. The author (RC) acknowledges the financial support from FAPERJ (no.E-26/171.754/2000, E-26/171.533/2002 and E-26/170.951/2006). MFAdaS  acknowledges the financial support from CNPq-Brazil, FINEP-Brazil (Ref. 2399/03), FAPERJ/UERJ (307935/2018-3) and from CAPES (CAPES-PRINT 41/2017).

\textcolor{black}{
\section{Appendix A}
%
The aether field equations, collecting the terms $c_2$, $c_{13}$ and $c_{14}$, are given by
\bqn
&&G^{aether}_{tt}=-\frac{1}{8A (a^2 B+1) B^2 r^2}\times\nb\\
&&[\,\, c_{13}\, (-4 a'^2 r^2 A^2 B^2+3 a^2 B'^2 r^2 A^2+3 a^2 A'^2 r^2 B^2+3 a^4 A'^2 r^2 B^3-\nb\\
&&2 r^2 A B^2 a^4 B' A'-8 a r^2 A B^2 A' a'-2 a^2 r^2 A B A' B'-8 a r^2 A^2 B a' B'-\nb\\
&&8 r^2 A B^3 a' A' a^3-8 r^2 A^2 B^2 a' a^3 B'-8 r^2 B^3 A^2 a'' a^3-4 a^2 r^2 A B^2 A''-\nb\\
&&4 r^2 A B^3 A'' a^4-4 r^2 A^2 B^2 a^4 B''-4 a^2 r^2 A^2 B B''-8 a r^2 A^2 B^2 a''-\nb\\
&&4 a^2 r^2 A^2 B^3 a'^2+3 a^4 r^2 A^2 B B'^2-16 r A^2 B^3 a^3 a'-8 a^4 r A^2 B^2 B'-\nb\\
&&8 a^2 r A^2 B B'-16 a r A^2 B^2 a'-8 a^4 A B^3 A' r-8 B^2 A r A' a^2+8 B^2 A^2 a^2+\nb\\
&&8 A^2 B^3 a^4) -\nb\\
&&\,\, c_{14}\, (8 a'^2 r^2 A^2 B^2-2 a^2 B'^2 r^2 A^2-6 a^2 A'^2 r^2 B^2-3 a^4 A'^2 r^2 B^3-\nb\\
&&2 B' A' r^2 A+8 A r A' B+4 A'' B r^2 A+2 r^2 A B^2 a^4 B' A'+8 a r^2 A B^2 A' a'+\nb\\
&&12 a r^2 A^2 B a' B'+8 r^2 A B^3 a' A' a^3+8 r^2 A^2 B^2 a' a^3 B'+8 r^2 B^3 A^2 a'' a^3+\nb\\
&&8 a^2 r^2 A B^2 A''+4 r^2 A B^3 A'' a^4+4 r^2 A^2 B^2 a^4 B''+4 a^2 r^2 A^2 B B''+\nb\\
&&8 a r^2 A^2 B^2 a''+4 a^2 r^2 A^2 B^3 a'^2-3 a^4 r^2 A^2 B B'^2+16 r A^2 B^3 a^3 a'+\nb\\
&&8 a^4 r A^2 B^2 B'+8 a^2 r A^2 B B'+16 a r A^2 B^2 a'+8 a^4 A B^3 A' r+\nb\\
&&16 B^2 A r A' a^2-3 A'^2 B r^2) -\nb\\
&&\,\, c_{2}\, (-32 a r A^2 B^2 a'-8 a^4 r A^2 B^2 B'-4 a^2 r^2 A^2 B^3 a'^2-8 a r^2 A B^2 A' a'-\nb\\
&&32 r A^2 B^3 a^3 a'-2 a^2 r^2 A B A' B'-8 r^2 A B^3 a' A' a^3-8 r^2 A^2 B^2 a' a^3 B'-\nb\\
&&2 r^2 A B^2 a^4 B' A'-8 a^2 r A^2 B B'-8 B^2 A r A' a^2-8 a^4 A B^3 A' r+\nb\\
&&3 a^4 r^2 A^2 B B'^2+3 a^2 A'^2 r^2 B^2-4 a'^2 r^2 A^2 B^2+3 a^2 B'^2 r^2 A^2+\nb\\
&&3 a^4 A'^2 r^2 B^3-8 a r^2 A^2 B a' B'-8 a r^2 A^2 B^2 a''-4 a^2 r^2 A^2 B B''-\nb\\
&&4 a^2 r^2 A B^2 A''-4 r^2 A^2 B^2 a^4 B''-4 r^2 A B^3 A'' a^4-\nb\\
&&8 r^2 B^3 A^2 a'' a^3) -\nb\\
&&\,\, (-8 a^2 r A^2 B B'+8 B^2 A^2 a^2-8 B^2 A^2+8 B A^2-8 B^3 A^2 a^2-\nb\\
&&8 r B' A^2)\,]=0,\nb\\
\lb{Gtt}
\eqn
\bqn
&&G^{aether}_{tr}=\frac{\epsilon}{4B^2 r^2 A (a^2 B+1) \sqrt{A (a^2 B+1)}}\times\nb\\
&&[\,\, c_{13}\, (4 A'^2 B^3 r^2 a^3+2 r^2 B^4 a^5 A'^2+2 B'^2 r^2 A^2 a-4 B^2 r^2 A^2 a''+\nb\\
&&2 A'^2 B^2 r^2 a-8 B^2 r A^2 a'-4 r^2 B^4 A^2 a'' a^4-2 A B^4 r^2 a^5 A''-\nb\\
&&2 B^3 r^2 A^2 a^5 B''-4 A r^2 B^3 a^3 A''-2 B^2 r^2 A a A''-2 B r^2 A^2 a B''-\nb\\
&&4 B^2 r^2 A^2 a^3 B''-8 B^3 r^2 A^2 a^2 a''-8 B^2 r A^2 a^3 B'-16 B^3 r A^2 a^2 a'-\nb\\
&&4 B r A^2 a B'-8 B^4 A^2 r a^4 a'-4 B^3 r A^2 a^5 B'-4 A B^4 a^5 A' r+\nb\\
&&2 B^2 r^2 A^2 a^5 B'^2-8 r A B^3 a^3 A'-4 r A B^2 a A'-2 B^2 r^2 A A' a'-\nb\\
&&2 B r^2 A^2 a' B'+4 B r^2 A^2 a^3 B'^2-4 B^2 r^2 A^2 a' B' a^2-2 A B^4 r^2 A' a^4 a'-\nb\\
&&2 B^3 r^2 A^2 a^4 a' B'-4 A' B^3 r^2 A a' a^2+8 B^2 A^2 a+16 B^3 A^2 a^3+\nb\\
&&8 B^4 A^2 a^5) +\nb\\
&& \,\, c_{14}\,(-4 A'^2 B^3 r^2 a^3-2 r^2 B^4 a^5 A'^2-2 A'^2 B^2 r^2 a+4 B^3 r^2 A^2 a a'^2+\nb\\
&&4 r^2 B^4 A^2 a'' a^4+2 A B^4 r^2 a^5 A''+2 B^3 r^2 A^2 a^5 B''+4 A r^2 B^3 a^3 A''+\nb\\
&&2 B^2 r^2 A a A''+2 B^2 r^2 A^2 a^3 B''+4 B^3 r^2 A^2 a^2 a''+4 B^2 r A^2 a^3 B'+\nb\\
&&8 B^3 r A^2 a^2 a'+8 B^4 A^2 r a^4 a'+4 B^3 r A^2 a^5 B'+4 A B^4 a^5 A' r-\nb\\
&&2 B^2 r^2 A^2 a^5 B'^2+8 r A B^3 a^3 A'+4 r A B^2 a A'-B r^2 A^2 a^3 B'^2+\nb\\
&&6 B^2 r^2 A^2 a' B' a^2+2 A B^4 r^2 A' a^4 a'+2 B^3 r^2 A^2 a^4 a' B'+\nb\\
&&2 A' B^3 r^2 A a' a^2-B r^2 A a A' B'-B^2 r^2 A A' a^3 B') +\nb\\
&& \,\, c_{2}\,(4 A'^2 B^3 r^2 a^3+2 r^2 B^4 a^5 A'^2+2 B'^2 r^2 A^2 a-4 B^2 r^2 A^2 a''+\nb\\
&&2 A'^2 B^2 r^2 a-8 B^2 r A^2 a'-4 r^2 B^4 A^2 a'' a^4-2 A B^4 r^2 a^5 A''-\nb\\
&&2 B^3 r^2 A^2 a^5 B''-4 A r^2 B^3 a^3 A''-2 B^2 r^2 A a A''-2 B r^2 A^2 a B''-\nb\\
&&4 B^2 r^2 A^2 a^3 B''-8 B^3 r^2 A^2 a^2 a''-16 B^3 r A^2 a^2 a'-8 B^4 A^2 r a^4 a'+\nb\\
&&2 B^2 r^2 A^2 a^5 B'^2-2 B^2 r^2 A A' a'-2 B r^2 A^2 a' B'+4 B r^2 A^2 a^3 B'^2-\nb\\
&&4 B^2 r^2 A^2 a' B' a^2-2 A B^4 r^2 A' a^4 a'-2 B^3 r^2 A^2 a^4 a' B'-4 A' B^3 r^2 A a' a^2+\nb\\
&&8 B^2 A^2 a+16 B^3 A^2 a^3+8 B^4 A^2 a^5)=0,
\lb{Gtr}
\eqn
\bqn
&&G^{aether}_{rt}=0,
\eqn
\bqn
&&G^{aether}_{rr}=-\frac{1}{8A^2 B r^2 (a^2 B+1)}\times\nb\\
&&[\,\, c_{13}\, (a^4 r^2 A^2 B B'^2+4 a^2 r^2 A^2 B^3 a'^2+4 a r^2 A^2 B a' B'+4 r^2 A B^3 a' A' a^3+\nb\\
&&4 r^2 A^2 B^2 a' a^3 B'+2 r^2 A B^2 a^4 B' A'+4 a r^2 A B^2 A' a'+2 a^2 r^2 A B A' B'+\nb\\
&&8 B^2 A^2 a^2+8 A^2 B^3 a^4+4 a'^2 r^2 A^2 B^2+a^2 B'^2 r^2 A^2+a^2 A'^2 r^2 B^2+\nb\\
&&a^4 A'^2 r^2 B^3) -\nb\\
&&\,\, c_{14}\, (-4 a^2 r^2 A^2 B^3 a'^2-a^4 r^2 A^2 B B'^2-2 a^2 r^2 A B A' B'-4 r^2 A B^3 a' A' a^3-\nb\\
&&2 r^2 A B^2 a^4 B' A'-4 r^2 A^2 B^2 a' a^3 B'-4 a r^2 A B^2 A' a'-A'^2 B r^2-\nb\\
&&a^4 A'^2 r^2 B^3-2 a^2 A'^2 r^2 B^2) -\nb\\
&&\,\, c_{2}\, (16 a r A^2 B^2 a'+8 a^2 r A^2 B B'+8 B^2 A r A' a^2+8 a^4 A B^3 A' r+
\nb\\
&&a^4 r^2 A^2 B B'^2+16 r A^2 B^3 a^3 a'+8 a^4 r A^2 B^2 B'+4 a^2 r^2 A^2 B^3 a'^2+\nb\\
&&4 a r^2 A B^2 A' a'+4 a r^2 A^2 B a' B'+2 a^2 r^2 A B A' B'+4 r^2 A B^3 a' A' a^3+\nb\\
&&4 r^2 A^2 B^2 a' a^3 B'+2 r^2 A B^2 a^4 B' A'+16 B^2 A^2 a^2+16 A^2 B^3 a^4+\nb\\
&&a^2 A'^2 r^2 B^2+4 a'^2 r^2 A^2 B^2+a^2 B'^2 r^2 A^2+a^4 A'^2 r^2 B^3) -\nb\\
&& \,\,(8 B^2 A^2-8 B A^2+8 A^2 B^3 a^2-8 A r A' B-8 B^2 A r A' a^2-8 B^2 A^2 a^2)\,]=0,\nb\\
\lb{Grr}
\eqn
\bqn
&&G^{aether}_{\theta\theta}=-\frac{r}{8B^2 A r^2 (a^2 B+1)}\times\nb\\
&&[\,\, c_{13}\, (-A^2 r B a^4 B'^2-4 A^2 r B^3 a^2 a'^2-4 a r A^2 B a' B'-4 A r B^3 a^3 a' A'-\nb\\
&&4 A^2 r B^2 a^3 a' B'-2 B' A' r A a^2 B-2 A r B^2 a^4 B' A'-4 A r B^2 a a' A'+\nb\\
&&4 A' B^2 A a^2+4 B^3 A A' a^4-r B^3 A'^2 a^4-A'^2 B^2 r a^2+16 a A^2 B^2 a'+\nb\\
&&4 B' A^2 a^2 B+16 A^2 B^3 a^3 a'+4 A^2 B^2 a^4 B'-4 a'^2 r A^2 B^2-a^2 B'^2 r A^2) -\nb\\
&&\,\, c_{14}\, (A'^2 B r+4 A^2 r B^3 a^2 a'^2+A^2 r B a^4 B'^2+4 A r B^3 a^3 a' A'+\nb\\
&&4 A r B^2 a a' A'+4 A^2 r B^2 a^3 a' B'+2 A r B^2 a^4 B' A'+2 B' A' r A a^2 B+\nb\\
&&r B^3 A'^2 a^4+2 A'^2 B^2 r a^2)-\nb\\
&&\,\, c_{2}\, (8 a r A^2 B^2 a''+4 a^2 r A^2 B B''+4 A'' B^2 r A a^2+4 A^2 r B^2 a^4 B''+\nb\\
&&4 A r B^3 a^4 A''+8 A^2 r B^3 a'' a^3-3 A^2 r B a^4 B'^2+4 A^2 r B^3 a^2 a'^2+\nb\\
&&8 A r B^2 a a' A'+8 a r A^2 B a' B'+2 B' A' r A a^2 B+8 A r B^3 a^3 a' A'+\nb\\
&&8 A^2 r B^2 a^3 a' B'+2 A r B^2 a^4 B' A'+32 a A^2 B^2 a'+8 B' A^2 a^2 B+\nb\\
&&8 A' B^2 A a^2+8 B^3 A A' a^4+32 A^2 B^3 a^3 a'+8 A^2 B^2 a^4 B'-3 A'^2 B^2 r a^2+\nb\\
&&4 a'^2 r A^2 B^2-3 a^2 B'^2 r A^2-3 r B^3 A'^2 a^4) -\nb\\
&&\,\, (4 B' A^2 a^2 B-4 A' B^2 A a^2+2 A'^2 B^2 r a^2-4 A'' B r A+2 B' A' r A+\nb\\
&&4 B' A^2+2 B' A' r A a^2 B-4 A' B A-4 A'' B^2 r A a^2+2 A'^2 B r)\,]=0,
\lb{Gthetatheta}
\eqn
\bq
G^{aether}_{\phi\phi}=G^{aether}_{\theta\theta} \sin^2 \theta,
\lb{Gphiphi}
\eq
where $G^{aether}_{\mu\nu}=T^{aether}_{\mu\nu}$ and
the symbol prime denotes the differentiation with respect to $r$. We can notice here that when $c_{13}=0$, $c_{14}=0$ and $c_2=0$ we obtain the same field equations of the GR.
\clearpage
\section{Appendix B}
%
The aether field equations, collecting the terms $c_{123}$, $c_{14}$ and $c_2$, are given by
\bqn
&&G^{aether}_{tt}=\frac{1}{8A (a^2 B+1) B^2 r^2}\times\nb\\
&&[\,\, c_{14}\,(-8 a^3 r^2 A B^3 A' a'-8 a^3 r^2 A^2 B^2 a' B'-12 a r^2 A^2 B a' B'-\nb\\
&&2 a^4 r^2 A B^2 A' B'-8 a r^2 A B^2 A' a'-8 a^2 r A^2 B B'-4 a^2 r^2 A^2 B^3 a'^2-\nb\\
&&4 A^2 r^2 B a^2 B''-8 a^2 r^2 A B^2 A''-8 B^2 r^2 A^2 a a''-4 a^4 r^2 A^2 B^2 B''-\nb\\
&&4 a^4 r^2 A B^3 A''-8 a^3 r^2 A^2 B^3 a''+3 a^4 r^2 A^2 B B'^2-8 a^4 r A^2 B^2 B'-\nb\\
&&8 a^4 A B^3 A' r-16 r A^2 B^3 a^3 a'-16 a r A^2 B^2 a'-16 B^2 A r A' a^2+\nb\\
&&2 B' A' r^2 A-8 A r A' B+3 r^2 B^3 a^4 A'^2+6 a^2 A'^2 r^2 B^2-8 a'^2 r^2 A^2 B^2+\nb\\
&&2 a^2 B'^2 r^2 A^2-4 A'' B r^2 A+3 A'^2 B r^2) +\nb\\
&&\,\, c_{2}\, (16 a r A^2 B^2 a'+16 r A^2 B^3 a^3 a'+8 A^2 B^3 a^4+8 B^2 A^2 a^2) +\nb\\
&&\,\,  c_{123}\,(8 a^3 r^2 A B^3 A' a'+8 a^3 r^2 A^2 B^2 a' B'-8 B^2 A^2 a^2+8 a r^2 A^2 B a' B'+\nb\\
&&2 a^4 r^2 A B^2 A' B'+8 a r^2 A B^2 A' a'+2 a^2 r^2 A B A' B'+8 a^2 r A^2 B B'+\nb\\
&&4 a^2 r^2 A^2 B^3 a'^2+4 A^2 r^2 B a^2 B''+4 a^2 r^2 A B^2 A''+8 B^2 r^2 A^2 a a''+\nb\\
&&4 a^4 r^2 A^2 B^2 B''+4 a^4 r^2 A B^3 A''+8 a^3 r^2 A^2 B^3 a''-3 a^4 r^2 A^2 B B'^2+\nb\\
&&8 a^4 r A^2 B^2 B'+8 a^4 A B^3 A' r+16 r A^2 B^3 a^3 a'+16 a r A^2 B^2 a'+\nb\\
&&8 B^2 A r A' a^2-3 r^2 B^3 a^4 A'^2-3 a^2 A'^2 r^2 B^2+4 a'^2 r^2 A^2 B^2-\nb\\
&&3 a^2 B'^2 r^2 A^2-8 A^2 B^3 a^4)+\nb\\
&&\,\, (8 a^2 r A^2 B B'-8 B^2 A^2 a^2+8 A^2 B^3 a^2-8 B A^2+8 B^2 A^2+8 r B' A^2)\,]=0\nb\\
\lb{Gttb}
\eqn
\bqn
&&G^{aether}_{tr}=\frac{\epsilon}{4B^2 r^2 A (a^2 B+1) \sqrt{A (a^2 B+1)}}\times\nb\\
&&[\,\, c_{14}\, (2 B^2 r^2 A^2 a^3 B''+4 B^3 r^2 A^2 a^2 a''+2 B^4 r^2 A a^5 A''+2 B^3 r^2 A^2 a^5 B''+\nb\\
&&4 B^4 r^2 A^2 a'' a^4+4 A r^2 B^3 a^3 A''+2 B^2 r^2 A a A''+4 B^3 A^2 r a^5 B'+\nb\\
&&4 B^3 r^2 A^2 a a'^2+4 B^2 r A^2 a^3 B'-B r^2 A^2 a^3 B'^2+8 B^3 A^2 r a^2 a'+\nb\\
&&8 B^4 A^2 r a^4 a'-2 B^2 r^2 A^2 a^5 B'^2+8 r A B^3 a^3 A'+4 r A B^2 a A'+\nb\\
&&4 a^5 A B^4 r A'+6 B^2 r^2 A^2 a' B' a^2+2 B^4 r^2 A A' a^4 a'+2 B^3 r^2 A^2 a^4 a' B'+\nb\\
&&2 A' B^3 r^2 A a' a^2-B r^2 A a A' B'-B^2 r^2 A A' a^3 B'-4 A'^2 B^3 r^2 a^3-\nb\\
&&2 A'^2 B^2 r^2 a-2 B^4 r^2 a^5 A'^2)+\nb\\
&&\,\, c_{2}\, (4 B^3 A^2 r a^5 B'+4 B r A^2 a B'+8 B^2 r A^2 a^3 B'+8 r A B^3 a^3 A'+\nb\\
&&4 r A B^2 a A'+4 a^5 A B^4 r A') +\nb\\
&& \,\, c_{123}\, (-4 B^2 r^2 A^2 a^3 B''-8 B^3 r^2 A^2 a^2 a''-2 B^4 r^2 A a^5 A''-2 B^3 r^2 A^2 a^5 B''-\nb\\
&&4 B^4 r^2 A^2 a'' a^4-4 A r^2 B^3 a^3 A''-2 B^2 r^2 A a A''-4 B^3 A^2 r a^5 B'-\nb\\
&&8 B^2 r A^2 a^3 B'+4 B r^2 A^2 a^3 B'^2-16 B^3 A^2 r a^2 a'-8 B^4 A^2 r a^4 a'+\nb\\
&&2 B^2 r^2 A^2 a^5 B'^2-8 r A B^3 a^3 A'-4 r A B^2 a A'-4 a^5 A B^4 r A'-4 B r A^2 a B'-\nb\\
&&4 B^2 r^2 A^2 a' B' a^2-2 B^4 r^2 A A' a^4 a'-2 B^3 r^2 A^2 a^4 a' B'-\nb\\
&&4 A' B^3 r^2 A a' a^2-8 B^2 r A^2 a'+2 B'^2 r^2 A^2 a-4 B^2 r^2 A^2 a''+\nb\\
&&4 A'^2 B^3 r^2 a^3+2 A'^2 B^2 r^2 a+2 B^4 r^2 a^5 A'^2+8 B^2 A^2 a+16 B^3 A^2 a^3+\nb\\
&&8 B^4 A^2 a^5-2 B^2 r^2 A A' a'-2 B r^2 A^2 a' B'-2 B r^2 A^2 a B'')\,]=0,
\lb{Gtrb}
\eqn
\bqn
&&G^{aether}_{rt}=0,
\eqn
\bqn
&&G^{aether}_{rr}=-\frac{r}{8A^2 B r^2 (a^2 B+1)}\times\nb\\
&&[\,\, c_{14}\, (-4 a^2 r^2 A^2 B^3 a'^2-a^4 r^2 A^2 B B'^2-2 a^4 r^2 A B^2 A' B'-\nb\\
&&4 a^3 r^2 A B^3 A' a'-4 a^3 r^2 A^2 B^2 a' B'-4 a r^2 A B^2 A' a'-2 a^2 r^2 A B A' B'-\nb\\
&&A'^2 B r^2-r^2 B^3 a^4 A'^2-2 a^2 A'^2 r^2 B^2) -\nb\\
&&\,\, c_{2}\,(16 a r A^2 B^2 a'+8 a^2 r A^2 B B'+8 B^2 A r A' a^2+8 a^4 A B^3 A' r+\nb\\
&&16 r A^2 B^3 a^3 a'+8 a^4 r A^2 B^2 B'+8 B^2 A^2 a^2+8 A^2 B^3 a^4) -\nb\\
&&\,\, c_{123}\, (4 a^2 r^2 A^2 B^3 a'^2+a^4 r^2 A^2 B B'^2+4 a r^2 A B^2 A' a'+2 a^2 r^2 A B A' B'+\nb\\
&&4 a r^2 A^2 B a' B'+4 a^3 r^2 A B^3 A' a'+4 a^3 r^2 A^2 B^2 a' B'+2 a^4 r^2 A B^2 A' B'+\nb\\
&&8 B^2 A^2 a^2+8 A^2 B^3 a^4+4 a'^2 r^2 A^2 B^2+a^2 B'^2 r^2 A^2+a^2 A'^2 r^2 B^2+\nb\\
&&r^2 B^3 a^4 A'^2)-\nb\\
&& (-8 A r A' B-8 B^2 A^2 a^2-8 B A^2+8 B^2 A^2-8 B^2 A r A' a^2+8 A^2 B^3 a^2)\,]=0,\nb\\
\lb{Grrb}
\eqn
\bqn
&&G^{aether}_{\theta\theta}=-\frac{r}{8B^2 A^2 (a^2 B+1)}\times\nb\\
&&[\,\, c_{14}\, (A'^2 B r+4 r A^2 B^2 a' a^3 B'+2 A r B^2 a^4 B' A'+2 B' A' r A a^2 B+\nb\\
&&4 A r B^3 a^3 a' A'+4 A r B^2 a a' A'+4 r A^2 B^3 a^2 a'^2+r A^2 B a^4 B'^2+\nb\\
&&r B^3 A'^2 a^4+2 A'^2 B^2 r a^2) -\nb\\
&& \,\, c_{2}\, (12 A r B^2 a a' A'+12 a r A^2 B a' B'+4 B' A' r A a^2 B+12 A r B^3 a^3 a' A'+\nb\\
&&12 r A^2 B^2 a' a^3 B'+4 A r B^2 a^4 B' A'+8 a r A^2 B^2 a''+4 A r B^3 a^4 A''+\nb\\
&&4 r A^2 B^2 a^4 B''+8 r A^2 B^3 a'' a^3-2 r A^2 B a^4 B'^2+8 r A^2 B^3 a^2 a'^2+\nb\\
&&4 a^2 r A^2 B B''+4 A'' B^2 r A a^2+16 a A^2 B^2 a'-2 A'^2 B^2 r a^2+8 a'^2 r A^2 B^2-\nb\\
&&2 a^2 B'^2 r A^2-2 r B^3 A'^2 a^4+4 B' A^2 a^2 B+4 A' B^2 A a^2+4 B^3 A A' a^4+\nb\\
&&16 A^2 B^3 a^3 a'+4 A^2 B^2 a^4 B')-\nb\\
&&\,\, c_{123}\, (-2 B' A' r A a^2 B-2 A r B^2 a^4 B' A'-4 A r B^2 a a' A'-4 a r A^2 B a' B'-\nb\\
&&4 A r B^3 a^3 a' A'-4 r A^2 B^2 a' a^3 B'-r A^2 B a^4 B'^2-4 r A^2 B^3 a^2 a'^2+\nb\\
&&4 A' B^2 A a^2+4 B^3 A A' a^4-r B^3 A'^2 a^4-A'^2 B^2 r a^2+16 A^2 B^3 a^3 a'+\nb\\
&&4 A^2 B^2 a^4 B'+16 a A^2 B^2 a'+4 B' A^2 a^2 B-4 a'^2 r A^2 B^2-a^2 B'^2 r A^2) -\nb\\
&&\,\,(4 B' A^2 a^2 B-4 A' B^2 A a^2+2 A'^2 B^2 r a^2-4 A'' B r A+2 B' A' r A+\nb\\
&&2 A'^2 B r-4 A' B A+2 B' A' r A a^2 B+4 B' A^2-4 A'' B^2 r A a^2)\,]=0,
\lb{Gthetathetab}
\eqn
\bq
G^{aether}_{\phi\phi}=G^{aether}_{\theta\theta} \sin^2 \theta.
\lb{Gphiphib}
\eq
}

\section{References}

\end{document}